\documentclass[conference,compsoc]{IEEEtran}

\usepackage[utf8]{inputenc}
\usepackage[T1]{fontenc}
\usepackage{microtype}

\usepackage[pdftex]{graphicx}
\usepackage{tikz}
\usepackage[most]{tcolorbox}

\newcommand{\keyfinding}[1]{
    \begin{tcolorbox}[
        colback=gray!10,
        colframe=black,
        boxrule=0.8pt,
        title=Key Finding,
        fonttitle=\bfseries,
    ]
    #1
    \end{tcolorbox}
}

\usepackage{color}

\usepackage{tabularx}
\usepackage{booktabs}
\usepackage{threeparttable}
\usepackage{wasysym} 

\usepackage{enumitem}

\usepackage{xurl}

\usepackage[capitalise]{cleveref}

\usepackage{acronym}

\usepackage{fontawesome5}
\newcommand*\circempty{%
  {\fontsize{4.0pt}{4.0pt}\faCircle[regular]}}%\tiny\faCircle[regular]}
\newcommand*\circhalf{%
  {\fontsize{4.0pt}{4.0pt}\faAdjust}}
\newcommand*\circfull{%
  {\fontsize{4.0pt}{4.0pt}\faCircle}}

\begin{document}

\title{SoK: Understanding Anti-Forensics Concepts and Research Practices Across Forensic Subdomains}

\author{
\IEEEauthorblockN{
Janine Schneider\IEEEauthorrefmark{1}, 
Florian Ramming\IEEEauthorrefmark{2}, 
Maximilian Eichhorn\IEEEauthorrefmark{2}, 
Gaston Pugliese\IEEEauthorrefmark{2}, 
Chris Hargreaves\IEEEauthorrefmark{3}, }
\IEEEauthorblockN{Jan Gruber\IEEEauthorrefmark{4}, 
Joschua Schilling\IEEEauthorrefmark{5}, 
Julian Geus\IEEEauthorrefmark{2}, 
Kevin Mayer\IEEEauthorrefmark{6}, 
Lea Uhlenbrock\IEEEauthorrefmark{2}, 
Lena Voigt\IEEEauthorrefmark{2} and 
Frank Breitinger\IEEEauthorrefmark{1}
}
\IEEEauthorblockA{\IEEEauthorrefmark{1}University of Augsburg}
\IEEEauthorblockA{\IEEEauthorrefmark{2}Friedrich-Alexander-Universität Erlangen-Nürnberg}
\IEEEauthorblockA{\IEEEauthorrefmark{3}Oxford University}
\IEEEauthorblockA{\IEEEauthorrefmark{4}KASTEL Security Research Labs, Karlsruhe Institute of Technology}
\IEEEauthorblockA{\IEEEauthorrefmark{5}CISPA Helmholtz Center for Information Security}
\IEEEauthorblockA{\IEEEauthorrefmark{6}Rosenheim Technical University of Applied Sciences}
}

\maketitle

\begin{abstract}
Anti-forensics includes a growing set of techniques designed to obstruct forensic analysis. While cybercriminals increasingly rely on these methods, they also help researchers identify and remedy weaknesses in forensic tools, advancing the overall robustness of digital forensics.
Despite repeated efforts to define it, anti-forensics remains vague and inconsistent in its use. It also poses ethical challenges regarding the appropriateness of research practices and the legitimacy of the field itself.
This article presents a systematic analysis of 123~publications on anti-forensics, combining qualitative and quantitative methods.
We quantify the main techniques and attack vectors, examine their occurrence in different digital forensic subdomains, and identify typical research methods, motivations, and applications.
This work also discusses what these findings mean for future research and proposes directions for building a more coherent and ethically grounded understanding of anti-forensics.
\end{abstract}

\section{Introduction}

Since the widespread adoption of computers, digital forensics has become an essential component of modern investigations.
Deriving reliable insights from digital traces has steadily grown in importance as more aspects of daily life, communication, and crime have shifted into the digital realm.
As forensic methods matured, so did awareness among cybercriminals that these methods could reveal and prove their activities.

Over time, cybercriminals began to experiment with ways to mislead, obstruct, or entirely prevent digital forensic analysis.
Techniques to hide traces, manipulate evidence, or degrade data gradually became part of their operational toolbox.
Collectively, these techniques are referred to as digital anti-forensics.
Anti-forensics, therefore, developed alongside digital forensics.
The study of anti-forensics thus originated from practical experience.
Over time, the scientific community began to examine anti-forensics not only as a criminal technique but as a subject worthy of structured investigation.

Eventually, it was recognized that anti-forensic methods can also be developed for constructive purposes.
In analogy to offensive security, controlled anti-forensics can be used to test the robustness of forensic tools, workflows, and processes.
Carefully designed anti-forensic techniques can reveal blind spots, challenge oversimplified models, and help improve the resilience of forensic processes.
Still, this dual-use character makes anti-forensics research a controversial topic.
The same techniques that can strengthen forensic practices may also empower cybercriminals, mirroring long-standing debates in offensive cybersecurity research.

Before such debates can be meaningfully conducted, an understanding of what constitutes anti-forensics is required; clarifying this is far from straightforward.
Anti-forensics is not a single, coherent category but a diverse collection of strategies, tools, and behaviors that differ significantly across the various digital forensic subdomains.
What constitutes an anti-forensic act in disk forensics may look entirely different in memory forensics or multimedia forensics.
These differences complicate attempts to define the phenomenon and to compare findings across studies.

There have been several attempts to investigate anti-forensics, ranging from reviews \cite{jain_anti-forensics_2014, gul_survey_2017, Taneja_antiforensics_multimedia_review} to a systematic literature review \cite{gonzalez_arias_systematic_2024} and taxonomy work \cite{conlan_anti-forensics_2016}.
These works provide valuable insights but remain limited in scope and perspective.
To date, no Systematization of Knowledge has been conducted that consolidates existing research, identifies gaps, and offers a structured understanding of anti-forensics as a cross-cutting phenomenon within digital forensics.

Our contribution is therefore to provide a structured overview of digital anti-forensics as a research area and to highlight how conducting such research differs across digital forensics subdomains.
We examine in which subdomains anti-forensics plays the greatest role and which attack vectors receive the most attention. 
Furthermore, we analyze how the term anti-forensics is used within these subdomains, how and for what purposes the research is conducted, and whether the ethical dilemmas associated with this type of work are addressed.

Our research questions are therefore:
\begin{enumerate}[label=RQ\arabic*]
	\item What is the current state of anti-forensics research, and where do research gaps remain? 
	\item How is anti-forensics research conducted across the different subdomains of digital forensics? 
	\item Which areas of digital forensics are most at risk of being targeted by anti-forensic methods? 
	\item What are the different anti-forensics use cases?
\end{enumerate}

We investigate these questions through a systematic analysis of 123~publications on anti-forensics, combining qualitative and quantitative methods.
Although this work is intended as a Systematization of Knowledge work, we followed the guidelines of Kitchenham et al. \cite{Kitchenham_guidelines} to ensure that our approach is as objective as possible and can be replicated.

\section{Background}

\emph{``Digital forensics is a branch of forensic science that focuses on identifying, acquiring, processing, analysing, and reporting on data stored electronically.'' \cite{Interpol}}

As digital systems are diverse in architecture, function, and data representation, the field has been divided into subdomains.
Common examples include data storage forensics, memory forensics, mobile forensics, multimedia forensics, IoT forensics, and cloud forensics.
Each subdomain requires tailored methods and tools, as the types of data available, the volatility of evidence, and the technical constraints differ substantially across environments.
The definition and boundaries of digital forensics subdomains remain a subject of ongoing debate. Different frameworks provide different perspectives: the NIST Tools and Techniques Catalog \cite{noauthor_computer_nodate} emphasizes practical categories of forensic tools, whereas the DFRWS EU ten-year review \cite{breitinger_dfrws_2024} offers a historical overview of research trends within a fixed period. Some approaches, such as SOLVE-IT \cite{hargreaves_solve-it_2025}, intentionally avoid rigid domain-based divisions, recognizing that forensic techniques often span multiple areas and can be reused across contexts. This diversity of perspectives highlights that there is no universal agreement on subdomain classification. For the purposes of this study, we adopt a pragmatic approach, focusing on subdomains that capture the most relevant areas for anti-forensic research, while acknowledging the fluid and overlapping nature of these categories.

\subsection{What Is Digital Anti-Forensics?}
\label{sec:af_definition}

The field of anti-forensics is broad in scope and methodology, includ tools and techniques applicable in many different forensic subdomains.
To encompass all of these aspects, several attempts have been made to define anti-forensics (see \Cref{tab:definitions}).

Early definitions emerged from practical problems that practitioners encountered in their work.
Peron and Legary~\cite{peron_digital_2010} are among the first to define it as any attempt to \emph{``limit the identification, collection, collation and validation of electronic data''} relevant to an investigator, dividing anti-forensic methods into destroy evidence, hide evidence, manipulate evidence, and prevent creation of evidence.

Grugq~\cite{the_grugq_art_2005} views the goal of anti-forensics as \emph{``reducing the quantity and quality of forensic evidence'' and activities to ``mitigate the effectiveness of forensic investigation''}.
The author also discusses the exploitation of software bugs in forensic tools and names data destruction, data hiding, and data contraception as the main anti-forensic techniques.

Liu and Brown~\cite{liu_bleeding-edge_2006} define it as the \emph{``application of the scientific method to digital media in order to invalidate factual information for judicial review''}.
They describe anti-forensics as a useful tool testing technique, discuss the risks of misuse, and name three main targets for anti-forensic attacks: data, tools, and the analyst.

Rogers~\cite{rogers_anti-forensics_2006} defines anti-forensics as \emph{``attempts to negatively effect the existence, amount and/or quality of evidence from a crime scene, or make the analysis and examination of evidence difficult or impossible to conduct''}.
The author also defines categories such as data hiding, artifact wiping, trail obfuscation, and attacks against the forensic process or tools.

Harris~\cite{harris_arriving_2006} is the first to address the anti-forensic problem scientifically.
The author deems former definitions problematic, e.g., the definition by Peron and Legary~\cite{peron_digital_2010} omits the evidence analysis; the definition by Grugq~\cite{the_grugq_art_2005} ignores the forensic process.
In consequence, Harris builds upon those and considers anti-forensics to be \emph{``any attempts to compromise the availability or usefulness of evidence to the forensics process''}.
Harris proposes a basic categorization of anti-forensic techniques, which includes destroying evidence, hiding evidence, eliminating sources of evidence, and counterfeiting evidence.

Other work further maps the forensic threat landscape and offers taxonomies to clarify the topic \cite{kessler_anti-forensics_2007, sremack_taxonomy_2007}.

Conlan et al. \cite{conlan_anti-forensics_2016} summarize previous definitions and state that \emph{``a majority of the definitions emphasize that anti-forensics can be identified by any attempts to alter, disrupt, negate, or in any way interfere with scientifically valid forensic investigations''}.
They present a refined categorization of anti-forensic tools, based on Rogers~\cite{rogers_anti-forensics_2006}, which includes steganography, encryption, data hiding, attacks against forensic tools and methods, trail obfuscation, and artifact wiping.

\begin{table*}[ht]
\caption{Summary of the most important definitions of anti-forensics.}
\centering
\begin{tabularx}{\textwidth}{l|X}
    \textbf{Authors} & \textbf{Definition}\\
    \midrule
    Peron and Legary (2005) \cite{peron_digital_2010} & Attempt to limit the identification, collection, collation and validation of electronic data. \\ 
    \midrule
    Grugq (2005) \cite{the_grugq_art_2005} & Reducing the quantity and quality of \textbf{forensic evidence}. \\ 
    \midrule
    Liu and Brown (2006) \cite{liu_bleeding-edge_2006} & Application of the scientific method to digital media in order to invalidate factual information for \textbf{judicial review}. \\ 
    \midrule
    Rogers (2006) \cite{rogers_anti-forensics_2006} & Attempts to negatively effect the existence, amount and/or quality of \textbf{evidence from a crime scene}, or make the analysis and examination of evidence difficult or impossible to conduct. \\ 
    \midrule
    Harris (2006) \cite{harris_arriving_2006} & Any attempts to compromise the availability or usefulness of \textbf{evidence to the forensics process}. \\
    \midrule
    Kessler (2007) \cite{kessler_anti-forensics_2007} & Set of tools, methods, and processes that hinder such analysis. \\
    \midrule
    Sremack and Antonov (2007) \cite{sremack_taxonomy_2007} & Any \textbf{activity that intentionally} aims to deceive or impede the forensic analysis. \\
    \midrule
    Conlan et al. (2016) \cite{conlan_anti-forensics_2016} & Any attempts to alter, disrupt, negate, or in any way interfere with scientifically valid \textbf{forensic investigations}.
\end{tabularx}
\label{tab:definitions}
\end{table*}

A closer look at existing definitions shows that the context in which specific methods and tools are used is crucial. Most definitions refer to digital evidence, not data in general, and data becomes evidence only within a legal framework, which is explicitly mentioned in several definitions \cite{liu_bleeding-edge_2006, rogers_anti-forensics_2006, harris_arriving_2006, conlan_anti-forensics_2016}.
Technologies themselves are neutral, i.e., they can be used for legitimate and criminal or destructive purposes. Consequently, another important aspect is that the method must be \emph{intentionally used} for anti-forensic purposes as defined by Sremack and Antonov \cite{sremack_taxonomy_2007}.

\section{Methodology}

To advance understanding of anti-forensics, we conducted a structured analysis of existing literature.
Our methodology is based on the approach proposed by Kitchenham \cite{Kitchenham_guidelines} and includes an initial literature search, search term engineering, a final literature search, defining inclusion and exclusion criteria, the paper selection, data collection, and data analysis. Some more details are discussed in the upcoming subsections.

\subsection{Literature Mapping and Search Term Engineering}

We first conducted an initial literature search to obtain a broad overview of the existing research, using search engines such as Google Scholar and search terms such as ``anti forensics'', ``anti forensic'', ``counter forensic'', and ``counter forensics''.
Our analysis showed that small variations in search terms can greatly influence retrieved results. Moreover, the terms used fail to capture the full range of publications on anti-forensic techniques.
This is because some authors do not explicitly label their work as anti- or counter-forensic, but instead use descriptions such as \emph{``prevent detection by forensic analysis'' \cite{huebner_data_2006}}.
When using search terms derived from existing taxonomies, such as Conlan et al. \cite{conlan_anti-forensics_2016}, many papers that were initially missed appear in the results, since the taxonomy provides keywords that are agnostic to the application.
Based on these findings, we iteratively refined our search terms, discussed them with an expert panel, tested them, and revised them. 

\subsection{Final Literature Search}

We proceeded to construct our dataset by conducting the main literature search for this study.
For this purpose, we used the following search string:

\begin{center}
 ("digital" OR "cyber" OR "computer") AND\\ ("forensics" OR "forensic") AND\\ ("anti" OR "counter" OR "[KEYWORD]")   
\end{center}

The first two parts of the search string establish the general context of digital forensics, while the third part addresses the observation that authors often use terms beyond the labels anti- or counter-forensics to describe their work.
To capture this diversity, we incorporated keywords from Conlan's taxonomy \cite{conlan_anti-forensics_2016} to complement the search string.
To explore the literature, we queried the scientific databases ScienceDirect, IEEE Xplore, and ACM Digital Library.
To narrow down the results (e.g. to papers from the computer science domain), we applied additional filters available in the search interfaces.
We did not limit the search to titles, abstracts, and keywords but conducted a full-text search.
Our search concluded in July 2024, capturing the most recent contributions up to the time of our literature search.
The initial dataset included 12,853~papers.

\subsection{Inclusion and Exclusion Criteria}

We defined inclusion and exclusion criteria to select papers relevant to our research questions. Papers with any of the following characteristics were excluded, even if they met the inclusion criteria:

\begin{itemize}
	\item The article is a duplicate
    \item The article is not in English
    \item The article was published before the year 2000 
	\item The article has not been cited at least once (for articles before 2023)
	\item The article is not accessible
	\item The article is not available as a PDF
	\item The article is not a full research paper
\end{itemize}

We do not expect to find relevant articles before the year 2000, as the field of anti-forensics research only began to take shape in the early 2000s.

For the inclusion criteria, we focused on papers that explicitly position their work within an anti-forensic context, analyze anti-forensic techniques, or examine their impact on digital investigations.
We define anti-forensics research as work that addresses \textbf{anti-forensic methods}, in which these methods represent the \textbf{primary focus} of the study and where this is communicated through \textbf{explicit language}.
For defining anti-forensics, we adopt the definition provided by Conlan et al. \cite{conlan_anti-forensics_2016}, as it summarizes many earlier formulations.
Being the primary focus means that anti-forensics is the main topic of the paper, and is not just mentioned briefly or as a side note in research that primarily addresses another area.
Containing explicit language means that a paper either uses the same terminology as the anti-forensic definition to describe the context of its results or employs synonyms that convey the same meaning and context.

\subsection{Final Paper Selection}

After applying the exclusion criteria, we conducted an initial screening of titles and abstracts to exclude publications that were out of scope.
We established intercoder agreement by independently screening a sample dataset of 100~papers and then discussing and comparing the classification results.
This process ensured consistency in the application of the criteria and increased the reliability of our subsequent full dataset coding.
This reduced the dataset to 1,680~papers.
For the remaining papers, we evaluated whether each paper met the predefined inclusion criteria, which resulted in 123~papers. 

\subsection{Data Collection}
We proceeded with the systematic extraction of data from the included papers.
To organize this process, we developed a codebook, implemented as a structured spreadsheet containing all selected papers.
Some codes, such as the forensic subdomains, were predefined and fixed, whereas others were generated inductively during the review of the papers.
The codebook was periodically reviewed and updated after coding a set number of papers.
This iterative approach allowed us to incorporate new insights and refine categories as needed.
To ensure consistency and a shared understanding among all collaborators, we held regular meetings to discuss coding decisions, resolve ambiguities, and update the categorization scheme accordingly.

\section{Results}

\subsection{Overview of Surveyed Works}
\label{sec:overview}

We analyzed a total of 123~papers in our study.
The publications varied in length, with an average of 11~pages per paper. 
Most papers focused on counter-anti-forensic approaches~(57), followed by studies exclusively investigating anti-forensic techniques~(36), and studies addressing both anti-forensic and counter-anti-forensic aspects~(30).

The majority of studies were situated in the domain of data storage~(62~papers), followed by multimedia~(16), overview~(11), mobile devices~(6), memory~(6), malware and network forensics (4 each), cloud forensics~(3) and IoT, email, vehicles, and GPU forensics (1 each).
Thirteen papers were classified as other forensic subdomains.
Among these, four papers addressed multiple subdomains simultaneously.

A notable number of papers did not explicitly use the term anti-forensics or any of its variants (29~papers).
Similarly, 24~papers did not discuss the anti-forensic potential of the concerned methods, highlighting that anti-forensics research is sometimes implicit rather than explicitly labeled.

The anti-forensic techniques studied varied widely.
The most frequently investigated attack was data falsification or manipulation (42~papers), followed by data hiding~(29), attacks against forensic tools or methods~(25), artifact wiping~(19), trail obfuscation~(14), encryption and steganography (each 8), and adversarial examples~(3).
Nineteen papers were categorized as other types of attack.
Thirty-four of these papers reported multiple anti-forensic types.

The most targeted technological aspect was file system content~(30), followed by file format~(24), the digital forensics process itself~(16), file system metadata~(15), operating system internals such as the Windows Registry~(10), application artifacts~(9), forensic tools~(8), memory~(8), network activity~(4), anti-forensic tools and firmware (each 3) and email~(1).
Ten papers had no specific target as they were exploring several attack vectors.

Most of the papers were written for digital forensic investigators, either to raise awareness of specific anti-forensic techniques or to present countermeasures~(109). This is followed by papers discussing anti-forensic approaches for surveillance protection~(7), tool testing for developers~(3), and two papers exploring the use of anti-forensics for intelligence purposes or to demonstrate to law professionals how technical methods could be used to create false alibis~(2).
Two papers did not specify their target audience.

Forty-five papers developed software, with 15 providing publicly available implementations. 
Additionally, five papers shared datasets.
Six papers proposed new taxonomies and/or definitions relevant to anti-forensics research.
Only four papers explicitly discussed ethical considerations.

The most common methodology used in anti-forensics research is experiments~(93), followed by literature reviews~(19), theoretical discussions~(9) and formal modeling~(7).
Five papers used mixed methodologies.

Summaries of our findings are depicted in the appendix: 
\Cref{tbl:overview:1} lists all papers exploring anti-forensics, 
\Cref{tbl:overview:2} highlights articles focusing on counter-anti-forensics, and \Cref{tbl:overview:3} presents articles including an anti-forensic approach \emph{and} a counter-anti-forensic approach (often mitigating the presented anti-forensic method).
Note that we use the term anti-forensics research for papers on anti-forensics and counter-anti-forensics.

\subsection{Usage of Anti-Forensics Terms}
\label{sec:AF_term}

Especially early works from the data storage subdomain \cite{jones_what_2004, huebner_data_2006, kwon_tool_2008} did neither use the term anti-forensics nor discuss anti-forensics with terms that are nowadays associated with it.
This may be because the term anti-forensics was more precisely defined in 2006 \cite{harris_arriving_2006}, bringing it into sharper focus within the digital forensics research community. 
In the subdomain of steganography, some early papers also did not use the term anti-forensics but discussed it with terms that are nowadays associated with anti-forensics \cite{castiglione_taking_2007, savoldi_data_2007}. 

In addition to the definitions discussed in \Cref{sec:af_definition}, further definitions for anti-forensics have been proposed by \cite{goh_plausibly-deniable_2011, dafale_sensor_2017, sencar_digital_2013}.
\cite{goh_plausibly-deniable_2011} address key aspects of anti-forensics by defining data tampering as a process that \emph{``involves deliberate tampering of the computer systems with the aim of altering or destroying data that contains incriminating information before or during a forensic examination is conducted''}. 
\cite{dafale_sensor_2017} presented another definition of anti-forensics adapted from \cite{sencar_digital_2013}: \emph{``The research field that challenges digital forensics and systematically explores its limitations against intelligent counterfeits is called [...] anti-forensics''}. 
In the context of analyzing different methods for detecting timestamp manipulations, 
\cite{oh_forensic_2024} defined the term anti-forensic resistance as \emph{``resistance to anti-forensic operations, such as deletion, tampering, and initialization that an attacker might perform to erase traces of timestamp manipulation''}. 

In most subdomains, the number of papers discussing anti-forensics and those explicitly using the term are identical, e.g., data storage (46/46), malware (4/4), cloud (3/3), GPU (1/1), IoT (1/1), and vehicle (1/1).
Precisely, as soon as anti-forensics is addressed, the term is also used explicitly, resulting in consistent terminology.
In contrast, in the subdomains of memory, multimedia, and network, anti-forensics is discussed without explicitly using the term, meaning it is described implicitly or using alternative expressions.

\keyfinding{
   In most digital forensics subdomains, the term anti-forensics is explicitly used. Only a few subdomains describe anti-forensics using other terms. One of these domains is multimedia forensics, which is one of the most important domains in the field of anti-forensics.
}

\subsection{Anti-Forensics Across Subdomains}
\label{sec:subdomains}

For our study, we defined a set of digital forensics subdomains to structure the analysis, which includes: Data Storage, Multimedia, Mobile, Memory, Network, Malware, Cloud, IoT, Email, GPU, and Vehicle, each focusing on specific types of data, devices, or investigative methods.
Additionally, the Overview subdomain captures papers spanning multiple areas, such as surveys, systematic literature reviews, and systematization of knowledge studies.
We now analyze these subdomains and their significance within the context of anti-forensics.

\begin{figure}[b]
	\centering
	\includegraphics[width=\linewidth]{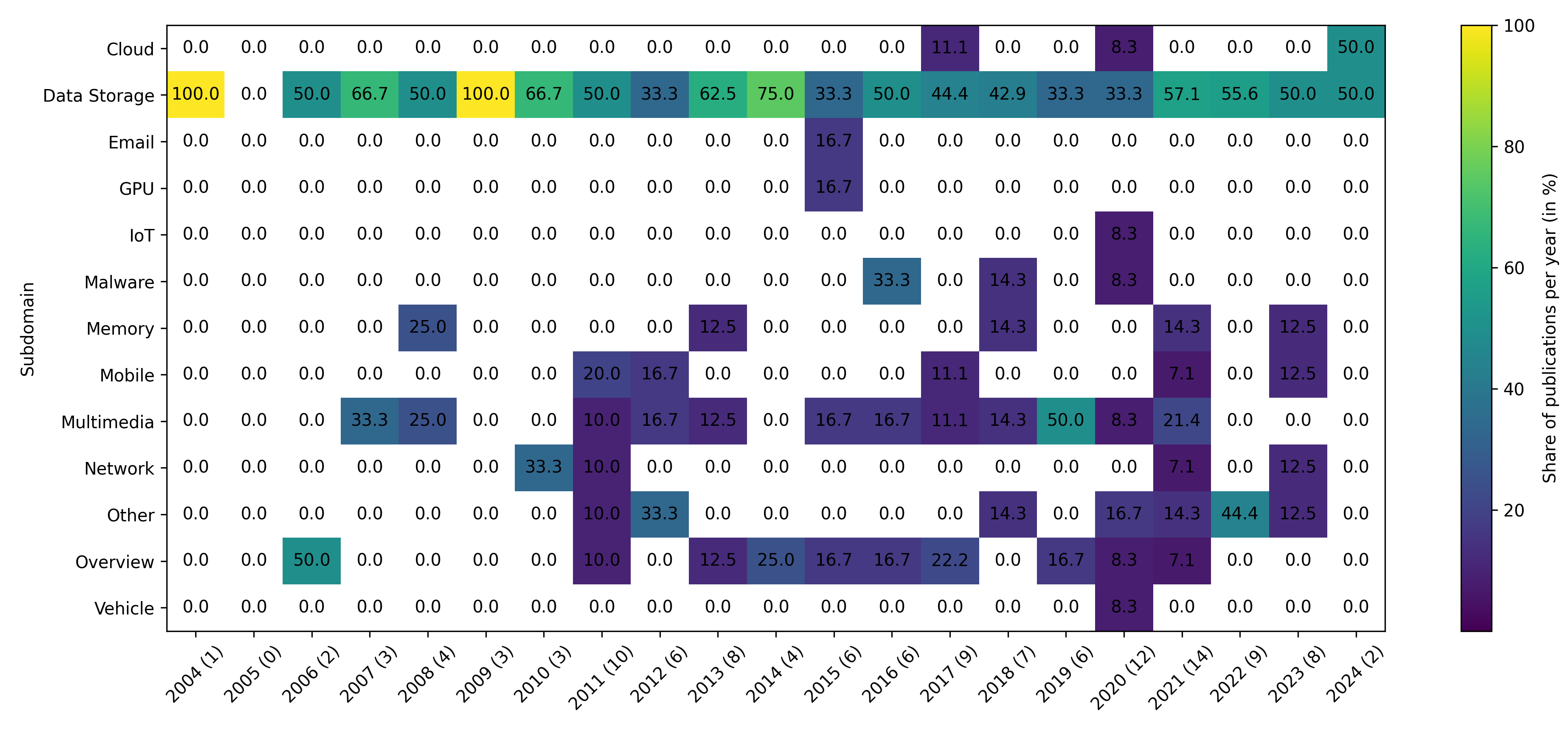}
    \caption{Distribution of anti-forensics papers (N=123) across digital forensics subdomains. Most research is conducted in data storage~(58) and multimedia~(16) subdomains. Since four papers could be assigned to multiple subdomains, this is reflected in the paper numbers per year. 
    % \todo[inline]{Todo here but in general: address double-counting due to 4 papers having multiple subdomains?}
    }
	\label{fig:subdomain}
\end{figure}

Data storage has been a focus of anti-forensics research since the 2000s and remains the predominant subdomain, as shown in \Cref{fig:subdomain}.
Furthermore, the proportion of data storage anti-forensic papers remains consistently high, with two peaks in 2004 and 2009 (see \Cref{fig:subdomain}) when this subdomain accounted for 100\% of our dataset. 
This high proportion of data storage is likely explained by the fact that computer forensics is the core area of digital forensics and its oldest subdomain.
The proportion of data storage papers in our dataset totals 48.1\%. 
However, anti-forensics research is becoming more diverse. 

The second key area, multimedia forensics, addresses image, video, and audio files and streams, covering manipulation and steganography, and comprises 12.4\% of our dataset.
Trail obfuscation, in the sense of concealing manipulation, also plays a major role in this subdomain. 
Accordingly, the multimedia subdomain involves a multitude of facets, leading to numerous potential avenues for attack.
However, it is important to note that deepfakes and other manipulation techniques in the field of fake news are not considered in our study, as the legal or investigative context is often missing or not explicitly stated. 
Otherwise, this area would probably be even larger. 
It can therefore be concluded that anti-forensics in the multimedia subdomain is not clearly enough defined, as motivation is decisive for classification as an anti-forensics technique according to the current definitions of anti-forensics, particularly in this area.

In other established subdomains, such as mobile, memory, network, and malware, there is a smaller but not negligible amount of anti-forensics research. 
However, it should be noted that the term anti-forensic is typically used in a different context in the malware subdomain, which is why many papers were excluded from the study due to a lack of legal or investigative context. 
In general, the term anti-forensics is used here primarily to describe malware that seeks to remain undetected for as long as possible to evade intrusion detection systems and antivirus programs.
Less commonly, malware is used to evade or prevent actual digital forensic analysis. 
The understanding of anti-forensics in the malware subdomain may be better captured by the term anti-incident response than by the term anti-forensics.
In the memory subdomain, research focuses primarily on data hiding and manipulation, as well as on attacks against tools/methods. 
The primary focus here is on hiding data and evading memory dump and analysis tools.

In addition to the established digital forensic subdomains, some domains are underrepresented, such as cloud, GPU, vehicle, and IoT. 
These areas are not only underrepresented in anti-forensics but also relatively new within digital forensics. Although our dataset includes only three cloud papers, they are noteworthy for illustrating the diversity of anti-forensic approaches in this subdomain.
This highlights that cloud forensics is a vast and complex field, and that anti-forensics research in this area must be equally comprehensive.
The complexity of the cloud environment likely makes it a challenging area for research, which could explain the limited number of dedicated anti-forensic studies.
In the subdomain of email forensics, only a single paper could be identified, indicating that this area may be too narrowly focused for extensive anti-forensic research.
Similarly, no papers were found in emerging areas such as drones, AI forensics (the digital forensic investigation of AI systems), and forensic AI (the use of AI for digital forensics).
These gaps suggest significant opportunities for future research.
Overall, existing digital forensics research involving AI predominantly focuses on applying AI tools, such as large language models, to support investigative tasks like drafting forensic reports, rather than addressing anti-forensic challenges.

A subset of our papers could not be assigned to a single traditional subdomain and was therefore categorized into the `Overview' or `Other' subdomains. Overview papers do not focus on a specific subdomain but instead provide a general synthesis or comprehensive survey of anti-forensics across digital forensics. The `Other' subdomain includes papers that do not fit established forensic domains, often covering interdisciplinary topics, novel technologies, or emerging areas.

\keyfinding{
   The anti-forensics landscape is dominated by the data storage and multimedia subdomains. Malware, in contrast, represents a distinct domain where the concept of anti-forensics may be more accurately described as anti-incident-response. Important but currently underrepresented areas are cloud, IoT, GPU, drones and AI.
}

\subsection{Anti-Forensic Use Cases}

This section outlines the intended audiences and practical applications of anti-forensics research, providing context for the purposes and scenarios in which such work is conducted.

Most of the papers we reviewed are aimed at forensic investigators, with authors typically framing their work around identifying and publicizing vulnerabilities.
Countermeasures against known anti-forensic methods or forensic challenges are, however, more commonly presented, representing the vast majority of publications.
A smaller subset of papers focuses on the development of anti-forensic techniques as tools for government authorities, intelligence agencies, and the military, or for private individuals such as activists, journalists, and minority groups seeking protection from persecution.
In this context, safeguarding against persecution emerges as the predominant motivation and constitutes the second-largest target audience.

Some papers, such as \cite{goh_plausibly-deniable_2011}, justify their work based on the right to protection against self-incrimination.
In certain legal systems, individuals may be compelled to provide decryption keys for their own data, making technologies that enable plausible deniability a valuable means of safeguarding personal or sensitive information (especially in autocratic systems).
To address this, \cite{goh_plausibly-deniable_2011} propose an encrypted system offering plausible deniability.
Specifically, they design an anti-forensic system leveraging the Trusted Platform Module (TPM) to impede forensic analysis, with the primary motivation being \emph{``to highlight the implications of such TPM-based approach in digital forensics for law enforcement agents''} and to protect sensitive data.
Another example in this context is \cite{khan_designing_2011}, which introduces a filesystem-based covert channel technique using cluster fragmentation to embed data, thereby achieving a two-fold form of plausible deniability.

Two papers justify their research by aiming to raise awareness among legal professionals. They explore how digital evidence can be manipulated to create false alibis \cite{castiglione_automated_2013, castiglione_forensic_2012}. The authors describe, inter alia, scenarios in which an insider could automate actions to implicate an unsuspecting victim and how the process of generating false alibis could itself be automated. The primary objective is to demonstrate to lawyers that digital evidence cannot be assumed to be inherently reliable and that court decisions should not rely solely on such evidence.

A relatively new phenomenon seems to be the development of anti-forensic tools for tool testing purposes. 
This shows a clear distinction between digital forensics and IT security, where it is common to take the attacker's perspective to test, harden, and make systems, software and processes more secure. Here, we see a clear gap in anti-forensics research and its application.

Only two papers do not specify the intended purpose of the proposed approach or the target audience for which it is intended. One of these papers presents several data hiding methods and an anti-forensic framework without addressing ethical considerations or clarifying for whom these techniques are intended.

\keyfinding{
   Anti-forensics research is predominantly aimed at digital forensic investigators and primarily serves to develop or evaluate countermeasures against known anti-forensic techniques. A smaller but significant portion of the research focuses on protecting sensitive data for government authorities, intelligence agencies, the military, or at-risk individuals, with plausible deniability playing a central role. Emerging approaches, such as the use of anti-forensic tools for testing purposes, highlight an underexplored opportunity to proactively harden systems and reveal a clear gap between digital forensics and established practices in IT security.
}

\subsection{Anti-Forensic Techniques}
\label{sec:af_types}

For distinguishing between different types of anti-forensics, we referred to the base categories proposed by Conlan et al. \cite{conlan_anti-forensics_2016} as stated in \Cref{sec:af_definition}.
To complement these well-established anti-forensic types and to include more recent concepts, we decided to extend the categorization with the terms adversarial example and others.
Additionally, we introduced a broader data falsification or manipulation category.

The most frequent type of anti-forensics is, by far, data falsification or manipulation. It is highly represented in the data storage and multimedia subdomain, mainly because research in these areas focuses on whether, and how, digital evidence can be convincingly falsified.
Timestamp manipulation represents a key topic in digital forensics, and a substantial body of work addresses the distinction between genuine and falsified timestamps \cite{galhuber_time_2021, oh_forensic_2024, jang_understanding_2016, mohamed_detection_2019}.
One factor contributing to the increased focus on manipulation in the data storage subdomain could also be the easier access to data in the field of traditional computer forensics, which accounts for a large share of papers in this subdomain. 
The manipulation of data in this context is, at least supposedly, less complex than in other areas.

As mentioned in \Cref{sec:subdomains}, the focus in the multimedia subdomain is often on whether the content of a multimedia file is credible, or whether an image can be attributed to a specific camera. 
Dafale and Naskar \cite{dafale_sensor_2017}, for instance, presented a photo response non-uniformity (PRNU) injection process, where the PRNU stemming from an original device $A$ is removed and replaced by the PRNU from another device $B$. Due to this, the correlation between the image and $B$ is higher than between the image and the original camera $A$ before tampering, thus effectively misleading forensic investigators. Similarly, Raj and Sankar \cite{raj_counter_2019} removed PRNUs from the original image and injected a new one. They found that the image quality is not negatively influenced by this and that the correlation of the image with the original camera is broken after anonymization. PRNUs can be falsified or manipulated, and in the multimedia subdomain, a large part of research is being conducted into how these processes can be detected.
At the same time, anti-forensic techniques related to multimedia, as well as the methods for evading detectors (i.e., essentially anti-counter-anti-forensics), are constantly being developed and, as is usual for machine learning, fine-tuned to maximum effectiveness.

Data hiding is the second largest area of anti-forensic types. 
It is particularly useful to be applied within highly structured data, such as file systems and databases, but data can also be hidden in file formats. 
In most cases, file system structures are less suitable for hiding large amounts of data.  Often, rarely used data structures in file systems, and smaller unused and special areas in file systems, are used to hide small amounts of data \cite{heeger_exhide_2021, kim_null_2022, gobel_revisiting_2019, martini_detecting_2008, hermon_forensic_2022, huebner_data_2006, carrier_different_2010, gobel_anti-forensics_2018}. 
There is only one approach where entire partitions are hidden \cite{schneider_ambiguous_2022}; however, these are primarily hidden from automated evaluation by forensic tools, but can be detected upon closer analysis of the memory.

Steganography, on the other hand, is better suited to effectively hide large amounts of data, such as images or larger texts. Nevertheless, steganography plays a rather minor role in anti-forensics research, as it is a traditional topic of multimedia forensics.
Steganography could be considered a subset of the broader data hiding category, as defined by Conlan et al. \cite{conlan_anti-forensics_2016}. However, we choose to distinguish it as a separate category due to its highly specialized nature and widespread use in multimedia settings.

The encryption of data usually also pursues the goal of hiding data, although masking would probably be a better term here. However, with the prevalence of encryption nowadays, especially in the mobile sector, encryption can hardly be described as an anti-forensic method in the sense of the definition described in \Cref{sec:af_definition} anymore.
Instead, it is more to be considered a challenge in digital forensics, which we will further elaborate in \Cref{sec:challenges}.

Other major areas of anti-forensic types include attacks against tools or methods, artifact wiping, and trail obfuscation. The first is probably the most interesting anti-forensic type, as it involves attacking the analysis or the digital forensic process itself. The goal is to prevent or slow down the analysis, generate false results, or even attack the analysis computer itself, as described in \cite{wundram_anti-forensics_2013}. The artifact wiping type is often determined by the need to remove traces of anti-forensic methods. Research in this area is therefore driven by the fact that this is a common sub-goal of anti-forensics.

Adversarial examples are not dealt with very often overall, as this topic is not usually discussed in the context of digital investigations. Here, we see a clear call to action, since machine learning classifiers are increasingly being used in digital forensics, for instance, to classify child sexual abuse material (CSAM).
The risk of adversarial examples, adversarial patches, and other more advanced methods influencing classification is greatly underestimated. 
Another example of the use of machine learning in digital forensics is described in \cite{nowroozi_employing_2023}, where the authors present a counter-anti-forensic method for adversarial attacks on neural networks that are used for analyzing network traffic. They trained a neural network specifically with genuine and malicious data to harden the neural network against adversarial attacks. 

The anti-forensic type other in our analysis included primarily overview papers, as they cover a broad and indeterminate range of subdomains and types of anti-forensics.

\keyfinding{
   The most prevalent type of anti-forensics is data falsification and manipulation. Encryption, although historically considered an anti-forensic technique, can no longer generally be classified as such. Attacks targeting forensic tools and processes hold substantial potential for disruption, including slowing down analysis, generating false results, or even compromising forensic systems. Adversarial examples remain highly underexplored despite their increasing relevance, particularly with the growing use of machine learning in digital forensics, representing a clear and underestimated risk that warrants further attention.
}

\subsection{Targets and Methods Used}

Before presenting the results, it is important to clarify the concept of a target, which is central to our analysis. In this study, a target refers to the object of both anti-forensic and counter-anti-forensic measures.
While anti-forensic techniques directly attack these targets, counter-anti-forensic methods aim to defend them against such attacks.
Accordingly, the term encompasses the objectives of both domains, covering both the attack and the protection of a target.

The two most prominent targets in our study are file system content and file formats.
Their prominence can be explained by the dominant role of data storage and multimedia in the analyzed papers.
Data storage targets include file system content, file system metadata, and operating system internals.
In contrast, multimedia targets are mainly file formats, as this category encompasses both manipulation of multimedia files and steganography.

Other identified targets include digital forensic processes, application artifacts, forensic tools, memory, network activity, anti-forensic tool usage, firmware, and email.

Anti-forensic tool usage refers to counter-anti-forensics papers that specifically aim to detect the use of anti-forensic tools.
Savoldi et al. \cite{savoldi_data_2007} study data hiding on SIM cards, while Zax and Adelstein \cite{zax_faust_2009} investigate residual artifacts left by popular steganography programs after uninstallation.
Oh et al. \cite{oh_forensic_2024} propose a machine-learning-based approach to detect traces of data wiping on a disk and to identify the tools and methods used for wiping.
This makes Oh et al. one of only three papers to apply machine learning for counter-anti-forensics.
Another study following this approach is Mohamed et al. \cite{mohamed_detection_2019}, which uses machine learning to detect NTFS timestamp tampering and thereby reduce the effort required for detection.
In contrast, Savoldi et al. \cite{savoldi_statistical_2012} employ a linear binary classifier to distinguish between wiped and standard disk sectors.
This represents a promising area of research that should be further pursued.
Machine learning for counter-anti-forensics remains underrepresented but offers significant potential for the efficient detection of anti-forensic tool usage.

Attacks on the digital forensic process represent one of the most widespread targets and appear across a variety of subdomains.
This highlights the importance of this target and the need to protect not just individual technologies but the forensic process as a whole against anti-forensic attacks.
Many papers addressing the digital forensic process and its protection focus on maintaining the integrity of the chain of custody.
Here, blockchain-based approaches dominate, aiming to safeguard the chain of custody from tampering \cite{martin_data_2022, ahmad_blockchain-based_2020, shahaab_preventing_2021}.
Notably, this approach is one of the few proactive counter-anti-forensics strategies.
The distinction between proactive and reactive counter-anti-forensic approaches is discussed in more detail in \Cref{sec:caf}.

Another paper that targets the digital forensic process is \cite{zubair_control_2022}.
where the authors perform a control-logic attack against industrial control systems (ICS) engineering software.
Such software may be used in digital forensic investigations to decompile and analyze ICS code.
The attack effectively renders the software unusable, thereby potentially hindering forensic analysis and disrupting the investigative process.

Several papers have also addressed forensic tools as a target.
As described in \Cref{sec:af_types}, the goal in this context is to prevent or slow down analyses, produce incorrect results, or even attack the analysis system itself.
Combined with attacks on the digital forensic process, this arguably represents the most critical attack surface, as the underlying technology becomes largely irrelevant if the analysis tool or process itself is compromised.
For example, a targeted attack on an automated, multifunctional tool can prevent the analysis of numerous devices and technologies.
Furthermore, the credibility of results obtained with forensic tools can be undermined if vulnerabilities are discovered, even if they are not actively exploited.
Hardening forensic tools and processes is therefore essential and should consider all possible measures to achieve this goal, including targeted anti-forensics research.
Two papers that specifically address this topic are \cite{schmitt_introducing_2018} and \cite{wundram_anti-forensics_2013}, the latter of which describes several attacks on forensic tools to raise awareness and for tool-testing purposes, and presents a taxonomy related to this subject.

In addition to the targets discussed before, we also want to highlight several particularly interesting targets and methods.
In the area of memory forensics, as well as memory as a target, the work by Palutke et al. \cite{palutke_styx_2018} is of particular note.
The authors propose Styx, a hardware‑based hypervisor rootkit that hides its memory artifacts in concealed memory regions.
They further discuss how this technique could be used as an undetectable forensic tool and therefore as a proactive counter‑anti‑forensics approach.
What initially appears detrimental to forensic investigations can thus be transformed into a specialized form of counter‑anti‑forensics.

Another interesting target and approach is presented in \cite{yu_covert_2015}.
In this paper, the authors discuss four examples illustrating how spam email can be used by criminals to hide information.

The paper by Franzen et al. \cite{franzen_randcompile_2023} is primarily a security paper but employs an anti-forensic technique to prevent malicious actors from automatically deriving kernel profile information (such as hardcoded comm values, the kernel's symbol table, ABI constraints, the order of structure fields, and pointer graphs).
As a result, the authors also hinder the fast and automated generation of kernel profiles for memory forensics tools such as Volatility.
They obfuscate this information by using string and pointer encryption, improved structure layout and parameter randomization, adding bogus parameters, and externalizing printk format strings at compile time for the Linux kernel.
In this case, anti-forensics becomes a security tool, as the goal is not to hide or obscure personal data but to protect technical information.

Another underrepresented area is surveillance cameras. Valente et al. \cite{valente_improving_2019} introduce visual challenges designed to make forging surveillance camera evidence more difficult. They examine different types of attacks and assess how effectively these attacks can be detected.
Bakas et al. \cite{bakas_detection_2021} propose a detection technique for inter-frame forgeries in surveillance videos.

A target that has received little research attention so far, but is becoming increasingly important due to the growing reliance on AI models such as large language models, is the GPU.
Only one of the analyzed papers addresses this topic.
Bellekens et al. \cite{bellekens_data_2015} extend existing methods for performing GPU memory forensics, discuss three anti-forensics techniques relevant to GPU memory forensics, and provide a methodology for conducting digital forensic investigations on GPUs.

\keyfinding{
   While some targets, such as memory, digital forensic processes, and forensic tools, receive considerable attention, many others, including surveillance cameras, are underrepresented. Several studies demonstrate that anti-forensics techniques can be repurposed for proactive counter-anti-forensic or security purposes, highlighting both the potential and the untapped research opportunities in these areas.
}

\subsection{Research Landscape}

\begin{figure}[b]
	\centering
	\includegraphics[width=\linewidth]{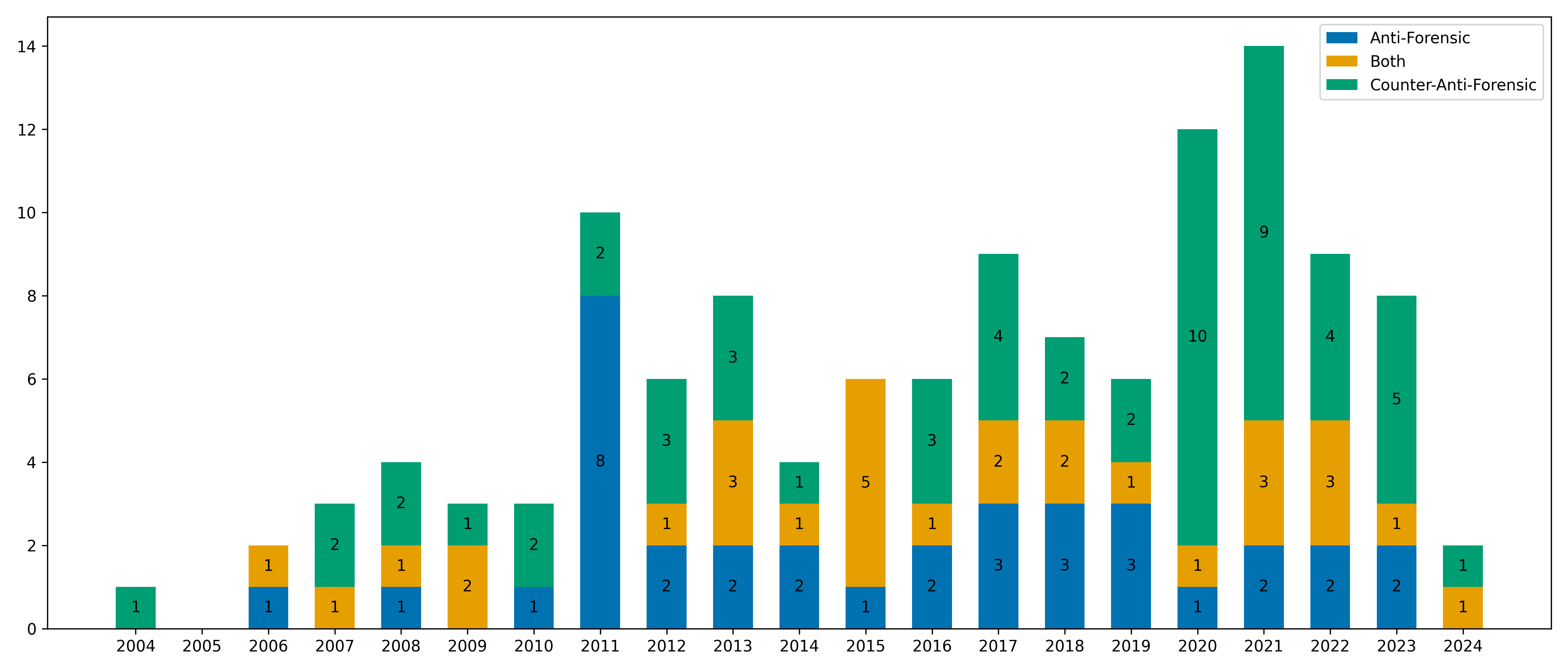}
    \caption{Distribution of papers (N=123) over publication years (2004--2024) that either focused on anti-forensics, counter-anti-forensics, or both. Most of the papers were published in 2021. 
    Overall, it is apparent that the percentage of pure anti-forensic papers is slowly declining over the years, while the percentage of counter-anti-forensic and both papers is increasing.
    }
	\label{fig:year}
\end{figure}

In terms of the type of anti-forensics research, counter-anti-forensics dominates across all subdomains.
Most studies focus on defense and the development of countermeasures rather than on attacker techniques, indicating that the field prioritizes protecting digital investigations.
This represents a positive trend (see \Cref{fig:year}), as it shows that research is increasingly aimed at strengthening forensic processes and mitigating potential threats.
Overall, the prevalence of counter-anti-forensic studies suggests that, while understanding attack strategies remains important, the primary emphasis is on safeguarding evidence and ensuring the reliability of digital investigations.
This can be viewed as both positive and negative. 
On the one hand, it dispels concerns that anti-forensics research is generating large-scale methods to undermine digital forensic processes, thereby enabling criminals to find new ways to evade law enforcement agencies, which is clearly not the case.
On the other hand, anti-forensics research has not yet achieved the same prominence within digital forensics as offensive security in the field of IT security.
These two aspects could be reconciled through appropriate rules, such as responsible reporting/disclosure, creating proof-of-concept solutions rather than full services, and involving the target audience and considering ethics. 
Both anti-forensics and digital forensics can still learn much from offensive security and its approaches in IT security.

One potential solution to this dilemma could be to address both the attacks and the countermeasures.
This is already the case in a non-negligible portion of the papers.
In these studies, besides the actual attack, authors present software tools and algorithms as countermeasures, alongside theoretical descriptions, recommendations, or artifacts and warning signals indicating the use of anti-forensic techniques.
Overall, the majority of papers %categorized as counter-anti-forensic or both 
focus on detecting traces of anti-forensic activity.

Further, we found that the predominant type of research is practical work and experimentation. 
Because digital forensics is highly practical, the same holds for anti-forensics, making it clearly an applied research field.

However, there are also other approaches, e.g., in  \cite{freiling_controlled_2018, schneider_tampering_2020, schneider_prudent_2022}. 
In these papers, a series of controlled experiments is conducted with students and professionals to investigate the difficulty of unnoticed data falsification. 
The authors experiment with browser and main memory traces and find that falsifying them unnoticed is difficult because each falsification action creates inconsistencies that must then be adjusted. 
A practical, empirical approach was therefore chosen here.

Nevertheless, as noted earlier, applied anti-forensic research increasingly attempts to formally define the field and develop models to describe anti-forensic techniques, countermeasures, and potential counter-anti-forensic processes.
For example, \cite{rekhis_hierarchical_2012} proposes a theoretical model of digital investigations that addresses anti-forensic attacks by emphasizing hierarchical visibility and the provability of anti-forensic actions.
In a series of papers, \cite{hasanabadi_memory-based_2021, hasanabadi_game-theoretic_2020, hasanabadi_survey_2020} explore the use of game theory to simulate anti-forensic attacks and identify optimal responses. They introduce a game-theoretic framework for modeling interactions between anti-forensic and counter-anti-forensic actions, allowing for the extension of a player's action space during simulations, with a particular focus on rootkits and anti-rootkits.
Another theoretical contribution is presented in \cite{liu_using_2012}, where the authors propose an adaptation of attack graphs that incorporates anti-forensic capabilities, enabling investigators to reconstruct attacks on vulnerable systems more effectively.

In addition to empirical and theoretical studies, a significant portion of anti-forensics research has focused on literature reviews to gain broader insights. Notably, temporal trends in these reviews reveal an evolution in focus over time. Early literature aimed at defining anti-forensics and establishing common terminology. Subsequently, research shifted toward comprehensive surveys that cataloged existing anti-forensic techniques \cite{al_fahdi_challenges_2013, jain_anti-forensics_2014, hausknecht_anti-computer_2017, gul_survey_2017, dahbur_anti-forensics_2011, al-mousa_general_2021, botas_counterfeiting_2015, karampidis_review_2018, adamu_conceptual_2020}, as noted by \cite{harris_arriving_2006}.
More recent reviews focus on classifying novel anti-forensic methods and technologies, as well as adapting concepts from other domains to the context of anti-forensics \cite{neale_fool_2023, neale_case_2022, hasanabadi_survey_2020}.
Three taxonomies have also been developed \cite{harris_arriving_2006, conlan_anti-forensics_2016, hasanabadi_survey_2020}.

\keyfinding{
   Anti-forensics research is predominantly applied and defense-oriented. Practical experimentation and software development are central, reflecting the field's strong emphasis on reproducibility and real-world applicability. Overall, the field exhibits a positive trend toward strengthening forensic processes, though it has yet to achieve the prominence and strategic integration seen in offensive security within IT security.
}

\subsection{Tools and Datasets}

\begin{figure}[b]
	\centering
	\includegraphics[width=\linewidth]{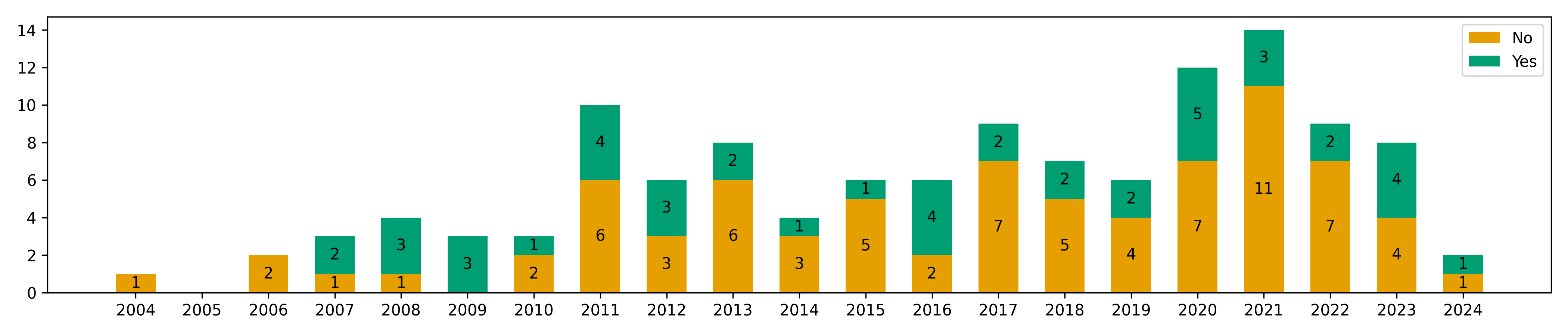}
    \caption{Number of papers (N=123) over publication years (2004--2024) that either did or did not develop software.}
	\label{fig:sw-developed}
\end{figure}

\begin{figure}[b]
	\centering
	\includegraphics[width=\linewidth]{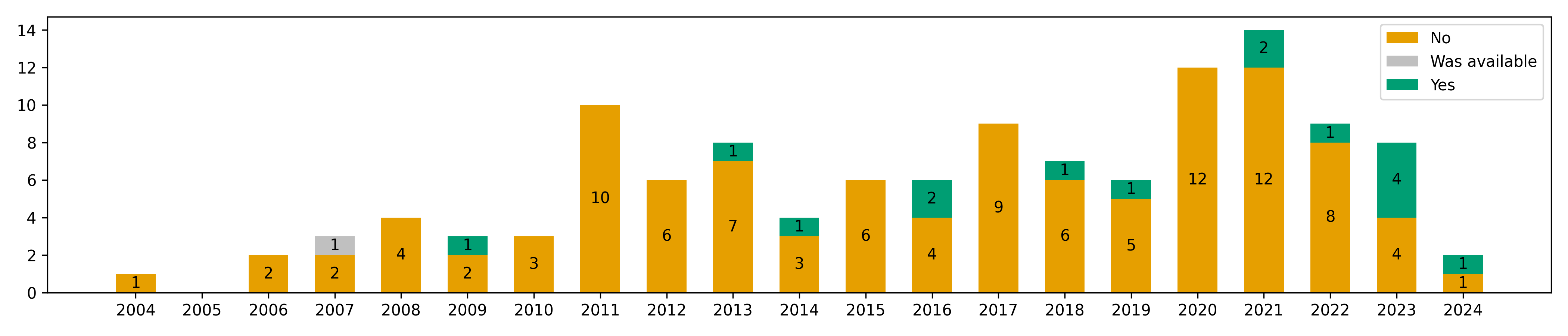}
    \caption{Number of papers (N=123) over publication years (2004--2024) that either did not provide software, whose provided software is no longer available, or provided software.} 
    %Overall, it can be seen that the proportion of papers with provided software has increased, especially in 2023.}
	\label{fig:sw-provided}
\end{figure}

Out of the 93~practical papers surveyed, 45 reported developing some form of software (\Cref{fig:sw-developed}).
Only 15 of these software projects were publicly accessible (\Cref{fig:sw-provided}).
This issue is not unique to anti-forensics research but reflects a broader problem in modern scientific practice.
It nevertheless provides only a partial explanation for why developed software is often not released.

The data storage domain also dominates software development.
In this area, 26~papers developed software, 10 of which made it publicly available, and 4 additionally provided a dataset.
This suggests that the focus on reproducible experimental work is particularly strong in this domain. 
In the cloud domain, one additional dataset was made available.
Lowetz et al. \cite{lowetz_anti-forensics_2024} show that anti-forensic techniques intended to hide user activities in cloud environments can be highly effective against analyses of forensic images taken from cloud virtual machines.
Their experiments are based on a dataset of 20~forensic images of Ubuntu virtual machines, which they also released for use by other researchers.

\keyfinding{
   Research software and datasets are being developed and partly made publicly available, especially in data storage forensics, but overall the number and accessibility remains limited.
}

\section{Discussion}

\subsection{What is Digital Counter-Anti-Forensics?}
\label{sec:caf}

The term counter-anti-forensics is not formally defined in the literature, but it generally refers to strategies and methods aimed at mitigating, neutralizing, or resisting anti-forensic techniques.
This can involve detecting traces of the usage of anti-forensic techniques, adapting tools and workflows to resist anti-forensic methods, and designing processes that remain robust even in the case of cybercriminals attempting to evade analysis.
In this sense, counter-anti-forensics is both reactive (i.e., identifying and recovering from anti-forensic actions) and proactive (i.e., anticipating potential countermeasures to ensure forensic outcomes remain trustworthy).

Most counter-anti-forensics papers focus on reactive methods, either detecting traces of anti-forensic activity or responding to previously proposed anti-forensic techniques.
In this context, the development of counter-anti-forensic methods is largely dependent on existing anti-forensics research and cannot proceed independently, as it primarily evolves as a response to anti-forensic techniques.
Nonetheless, there are also proactive approaches within counter-anti-forensics.

In addition to the blockchain-based approach for protecting the chain of custody, which was already presented, there are several other proactive strategies.
McDonald et al. \cite{mcdonald_enhanced_2017} propose an operating system design intended to limit tampering and anti-forensic activities.
Their approach introduces operating system functionality, such as component encryption, to prevent manipulation of selected files, for example, logs, and to make certain operating system functions resilient to tampering.
Hsu et al. \cite{hsu_digital_2011} suggest a formal model to secure the various steps of a forensic investigation. They provide formulas for each step to ensure integrity, confidentiality, and other security requirements.

The field of counter-anti-forensics can benefit greatly from approaches established in IT security, especially in terms of proactive measures to protect digital forensic processes.
Proactive strategies can be derived from techniques such as threat modeling and risk analysis, which allow potential vulnerabilities and attack vectors to be identified before an anti-forensic attack occurs.
In incident response, a related concept is forensic readiness, which focuses on preparing systems and processes to collect and preserve evidence for post-incident analysis. While forensic readiness involves proactive preparation, its primary goal is reactive: to facilitate investigation after an incident has occurred.
Proactive counter-anti-forensics, on the other hand, aims to actively prevent anti-forensic attacks from succeeding, rather than simply gathering information after the fact.
Overall, counter-anti-forensics combines both proactive and reactive approaches, integrating preventive measures with mechanisms for detection and response.

\keyfinding{
   There is currently no formal definition of counter-anti-forensics. It is important, however, to clearly distinguish it from forensic readiness. Proactive strategies based on IT security practices, such as threat modeling and risk analysis, offer a promising way to advance counter-anti-forensics.
}

\subsection{Anti-Forensics vs. Technical Challenges}
\label{sec:challenges}

As mentioned earlier, anti-forensics originated in the context of law enforcement practice.
In the early days of anti-forensics research, investigators faced various practical challenges in preserving and analyzing digital evidence.
These challenges were often addressed through techniques that deliberately altered, concealed, or obstructed evidence, which came to be recognized as the first anti-forensic methods.
Encryption plays a major role in early discussions on the subject of anti-forensics.
However, encryption can hardly be viewed as an anti-forensics method nowadays, considering the growing use of default encryption.
As discussed in \Cref{sec:af_definition}, technology itself is neutral.
The key factor lies in how the technology is used and the intentions behind its application.
The encryption activated by default on e.g. modern smartphones can therefore not be considered an anti-forensic method, even if this smartphone is part of an investigation.
Default encryption is therefore primarily a challenge for forensic analysis rather than an anti-forensic method.

Legal science, therefore, distinguishes between anti-forensics in a narrow sense and anti-forensics in a broader sense.
Anti-forensics, in a narrow sense, refers to measures deliberately taken to obstruct law enforcement or to counter specific investigative techniques.
Examples include dead man's switches triggered during police raids, the usage of encrypted messenger apps specifically developed for criminal purposes (such as the ANOM Messenger \cite{us_attorneys_office_southern_district_of_california_fbis_2021}), as well as the usage of deniable encryption to hide evidence.
A key aspect of these examples is the intent behind the use of these technologies.

A key area where anti-forensics, in the narrower sense, creates substantial obstacles for investigators is the use of cryptocurrencies in criminal contexts.
Cryptocurrencies are frequently used as a means of payment in criminal contexts, primarily due to the difficulty of reliably attributing transactions to specific individuals or purposes.
While most blockchain-based systems are transparent by design, they offer pseudonymity: transaction data is public, but the identity behind wallet addresses is not immediately apparent.
This pseudonymity, combined with the decentralized and borderless nature of cryptocurrencies, makes them attractive for criminal use cases.
To further evade attribution, criminals increasingly employ anti-forensic techniques designed to frustrate blockchain analysis \cite{deuber_sok_2022}.
Examples for such techniques are transaction mixers \cite{ziegeldorf_coinparty_2015} or CoinJoin protocols \cite{deuber_coinjoin_2021}.
Such measures are not limited to self-developed tools; commercial service providers actively offer transaction obfuscation as a professional service.
As a result, forensic blockchain analysis tools, designed to reconstruct transaction flows and attribute addresses to entities, have become a direct target of anti-forensic efforts.
The goal is to undermine traceability and render investigative techniques ineffective, particularly in cases involving money laundering, ransomware, or illicit marketplaces.

Anti-forensics in the broader sense refers to technologies that may complicate investigations, even if they are not intentionally used for anti-forensic purposes.
Examples of such technologies are end-to-end encryption or cloud services.
Such technologies could also be referred to as challenges for forensics and law enforcement agencies.

The challenges digital forensics has to face are usually a product of technological progress, while anti-forensic techniques are specifically designed to actively undermine forensic investigations: 
When a cloud storage provider deletes old data for economic reasons, it is a challenge for digital forensics.
When a hacker uses a tool that specifically deletes log files, it is anti-forensics.

Several of the analyzed papers should therefore be considered as addressing forensic challenges rather than anti-forensics.
For example, Casey~\cite{casey_growing_2011} discusses the growing impact of full-disk encryption on law enforcement authorities.
In contrast, deniable encryption is intentionally designed to conceal information and serves a very different purpose than everyday smartphone encryption.

Drawing a clear line between anti-forensics and forensic challenges can be difficult.
Consider, for instance, solid-state drives (SSDs), which pose significant forensic challenges due to wear-leveling behavior and other mechanisms.
These challenges are generally not considered anti-forensic, as SSDs themselves are not intentionally designed to obstruct investigations.
Similarly, the secure delete command alone is not usually regarded as an anti-forensic method.
However, using a secure erase tool on hard drives with the deliberate intent to thwart an investigation would be considered anti-forensic.

In some cases, challenges may be exploited for anti-forensic purposes.
For example, intentional data flooding or the use of proprietary software, e.g. in vehicle forensics, can create significant obstacles, even if this was not originally intended as an anti-forensic tool.
Conversely, the mere presence of large amounts of data or numerous devices, without malicious intent, cannot be classified as anti-forensics.
Context and intent are decisive.

Ultimately, the key factor in distinguishing between a forensic challenge and anti-forensics is the intention to evade law enforcement. Techniques or behaviors that arise naturally as technical challenges are not anti-forensic unless deliberately employed to obstruct investigations.

In this context, it may be useful to adopt the term counter‑forensic-effects, as used by Cannas et al. \cite{cannas2025jpeg}.
This term emphasizes the absence of malicious intent while still acknowledging that such effects can be deliberately leveraged as anti‑forensic techniques.

\keyfinding{
    Distinguishing between forensic challenges and anti-forensics depends primarily on intent. Technical obstacles are not inherently anti-forensic; they only become anti-forensic when deliberately used to hinder investigations. Concepts like counter-forensic-effects highlight that even unintentional technical effects can be exploited as anti-forensic techniques, emphasizing the importance of context and purpose.
}

\subsection{Ethical Awareness of Anti-Forensics Research}
\label{sec:ethics}

Despite anti-forensic techniques having potential real-world implications for law enforcement, ethical considerations appear to be no established aspect of anti-forensics literature, as only four papers in our sample addressed them explicitly, which is 3.25\%. The contexts for which ethics were discussed covered privacy \cite{odebade_mitigating_2017, alfosail_tor_2021} as well as motivations related to preventing forced disclosure of encryption keys in certain legal systems \cite{goh_plausibly-deniable_2011}, and demonstrating the feasibility of falsifying digital evidence \cite{castiglione_automated_2013}, 

While acknowledging that not discussing ethical implications explicitly does not indicate unethicalness per se, this may not apply to all research themes equally. Conducting experiments that showcase the effectiveness of anti-forensic techniques \cite{lowetz_anti-forensics_2024}, for instance, may seem counterproductive from certain points of views (e.g., putting the right tools in the wrong hands) but at the same time also help to raise awareness about existing imbalances in the arms race between anti-forensics and counter-anti-forensics, which should be in the interest of both practitioners and researchers in the digital forensics community. In other cases, where anti-forensic techniques have been proposed without addressing possible counter-anti-forensic measures or discussing ethical implications, the constructiveness may rightly be subject to discussions in the community.

\keyfinding{
    Ethical implications are largely unaddressed in the anti-forensics literature, which represents a difference to research areas where ethics considerations are more established, if not mandatory.
}

\section{Conclusion}
In this work, we have systematically studied 123~papers on anti-forensics published between 2004 and 2024.
Based on the insights gained, we identified several research gaps, outlined recommendations for future research, proposed more precise terms and definitions, and outlined potential learning from other IT security domains for digital forensics and anti-forensics research. 
Because only with precise definitions of terms can we conduct effective coordinated research, or to put it in Harris's words: 
\emph{``After all, if we cannot agree on how we should define the term and categorize the methods, how can we know an anti-forensic technique when we encounter it?''} \cite{harris_arriving_2006}. Returning to our initial research questions: 

\textbf{What is the current state of anti-forensics research, and where do research gaps remain?}
Overall, since the introduction of the first definitions for anti-forensics, the number of anti-forensic research papers has increased. 
The topic is therefore gaining in importance and is being viewed and researched from various angles.
The field of research is diverse, encompassing many forensic subdomains, applications, and objectives. 
Underrepresented but emerging areas include cloud, GPU, vehicle, IoT, drones, and especially AI, although we have not yet seen any anti-forensic research in the drone or AI subdomain. 
Another area that should be explored further is attacks on forensic tools and processes, as well as the use of machine learning for attack detection.

\textbf{How is anti-forensics research conducted across the different subdomains of digital forensics?}
In most digital forensics subdomains, the term anti-forensics is used according to the definition in~\cref{sec:af_definition}, with only a few early studies describing it in more descriptive terms. Research is predominantly conducted in an applied and empirical manner, with experimental studies forming the core methodology across all subdomains. Data storage and multimedia dominate anti-forensics research, likely due to their long-established history, the suitability of their data structures for manipulation and hiding, and comparatively easy access. Studies frequently focus on practical experiments, prototype development, and software creation, although the broader availability of developed software and datasets remains limited. Counter-anti-forensic research clearly prevails, particularly in detecting traces of anti-forensic tool usage and other defensive measures. Malware, mobile, and network subdomains see research that explores the implications of anti-forensic techniques on incident response and system analysis, while theoretical contributions, including game-theoretic models and adaptations of attack graphs, complement practical work by simulating attacks and evaluating countermeasures. Literature reviews and taxonomies provide additional structure, tracking temporal trends, new techniques, and cross-domain insights. Across all subdomains, ethical considerations are rarely addressed, and the focus remains overwhelmingly on defense rather than the development of offensive methods.

\textbf{Which areas of digital forensics are most at risk of being targeted by anti-forensic methods?}
The areas most at risk of anti-forensic attacks are data storage and multimedia, as they handle high-value digital evidence that can be manipulated, hidden, or falsified. Data storage is vulnerable due to the accessibility of storage media, making techniques like timestamp tampering and selective overwriting feasible. Multimedia is targeted for content authenticity attacks, such as PRNU manipulation or steganography. Emerging threats, including adversarial attacks on machine learning-based analyses and attacks against forensic tools, also put memory, network, and cloud forensics at increasing risk.

\textbf{What are the different use cases in which anti-forensic methods are applied?}
Anti-forensic methods are applied across several distinct use cases. The primary use case is the protection and defense of digital investigations, where counter-anti-forensic techniques are developed to detect, mitigate, or reverse the effects of anti-forensic activity. Another important use case is safeguarding sensitive or personal data, for example, through plausible deniability or encryption, especially in legal or autocratic contexts where individuals may be compelled to disclose information. Anti-forensics is also explored as a tool for raising awareness among legal professionals, illustrating the limits of relying solely on digital evidence and highlighting how false alibis can be created. Additionally, some research focuses on tool testing and system hardening, using anti-forensic techniques to evaluate the resilience of forensic tools and processes, similar to offensive security practices in IT. In a smaller set of cases, anti-forensic methods are developed for intelligence, government, or military applications, and occasionally for privacy protection by activists, journalists, or minority groups. Overall, these use cases range from defensive and educational purposes to controlled testing and the protection of sensitive information.

Our study has certain \textbf{limitations}: By using the legal or investigative context and the declared intent of the authors as inclusion criteria, we inevitably exclude many potentially interesting papers. However, this contextual framing is crucial for distinguishing anti-forensics from general challenges or dual-use tools. Additionally, our initial dataset highlighted the need for such limiting criteria to keep the scope of the study manageable. At the same time, this approach effectively filters out papers that discuss anti-forensic techniques without specifying the relevant context.

\textbf{Future research} could explore the extent to which the anti-forensic and counter-anti-forensic methods presented in the literature actually neutralize each other. At present, counter-anti-forensic studies significantly outnumber anti-forensic ones, but it remains unclear whether this surplus sufficiently addresses all anti-forensic techniques. A comprehensive approach to studying anti-forensics therefore requires examining both sides, attacks and countermeasures, within the same research framework, as demonstrated by some existing studies.

\section*{Acknowledgements}
G.~Pugliese was supported by the Bavarian Ministry of Science and Arts through the project ``Security in Everyday Digitization'' (ForDaySec).
J.~Gruber was funded by the DFG (German Research Foundation) through the Walter Benjamin Programme under grant number~570978749/GR~6850/1-1.
L.~Uhlenbrock and L.~Voigt were supported by the Research Training Group~2475 ``Cybercrime and Forensic Computing'' (grant number~393541319/GRK2475/1-2019).

\section*{Disclosure of AI-Assisted Writing Tools}
Several authors used ChatGPT, DeepL and Grammarly to help with tasks such as revising, condensing text, and addressing grammatical errors, typos, and awkward phrasing. All AI-generated suggestions were thoroughly reviewed and adjusted where needed to ensure they accurately reflected the authors' intended meaning before being incorporated into the paper.

\section*{Declaration of Interest}
The authors declare that they have no known competing financial interests or personal relationships that could have appeared to influence the work reported in this paper.

\bibliographystyle{IEEEtran}
\bibliography{bibliography}

@article{hargreaves_solve-it_2025,
	title = {{SOLVE}-{IT}: A proposed digital forensic knowledge base inspired by {MITRE} {ATT}\&{CK}},
	volume = {52},
	issn = {2666-2817},
	url = {https://www.sciencedirect.com/science/article/pii/S2666281725000034},
	doi = {10.1016/j.fsidi.2025.301864},
	series = {{DFRWS} {EU} 2025 - Selected Papers from the 12th Annual Digital Forensics Research Conference Europe},
	shorttitle = {{SOLVE}-{IT}},
	pages = {301864},
	journal = {Forensic Science International: Digital Investigation},
	shortjournal = {Forensic Science International: Digital Investigation},
	author = {Hargreaves, Christopher and van Beek, Harm and Casey, Eoghan},
	urldate = {2025-11-21},
	date = {2025-03-01},
    year = {2025},
}

@article{breitinger_dfrws_2024,
	title = {{DFRWS} {EU} 10-year review and future directions in Digital Forensic Research},
	volume = {48},
	issn = {26662817},
	url = {https://linkinghub.elsevier.com/retrieve/pii/S2666281723002044},
	doi = {10.1016/j.fsidi.2023.301685},
	pages = {301685},
	journal = {Forensic Science International: Digital Investigation},
	shortjournal = {Forensic Science International: Digital Investigation},
	author = {Breitinger, Frank and Hilgert, Jan-Niclas and Hargreaves, Christopher and Sheppard, John and Overdorf, Rebekah and Scanlon, Mark},
	urldate = {2025-11-21},
	date = {2024-03},
    year = {2024},
	langid = {english},
}

@online{noauthor_computer_nodate,
	title = {Computer Forensics Tools \& Techniques Catalog - Home},
	url = {https://toolcatalog.nist.gov/},
	urldate = {2025-11-21},
}

@book{sencar_digital_2013,
	location = {New York, {NY}},
	title = {Digital Image Forensics: There is More to a Picture than Meets the Eye},
	rights = {https://www.springernature.com/gp/researchers/text-and-data-mining},
	isbn = {978-1-4614-0756-0 978-1-4614-0757-7},
	url = {https://link.springer.com/10.1007/978-1-4614-0757-7},
	shorttitle = {Digital Image Forensics},
	publisher = {Springer},
	editor = {Sencar, Husrev Taha and Memon, Nasir},
	urldate = {2025-11-21},
	date = {2013},
    year = {2013},
	langid = {english},
	doi = {10.1007/978-1-4614-0757-7},
}

@inproceedings{deuber_coinjoin_2021,
	address = {Cham},
	title = {{CoinJoin} in the {Wild}},
	isbn = {978-3-030-88428-4},
	doi = {10.1007/978-3-030-88428-4_23},
	language = {en},
	booktitle = {Computer {Security} – {ESORICS} 2021},
	publisher = {Springer International Publishing},
	author = {Deuber, Dominic and Schröder, Dominique},
	editor = {Bertino, Elisa and Shulman, Haya and Waidner, Michael},
	year = {2021},
	pages = {461--480},
}

@inproceedings{ziegeldorf_coinparty_2015,
	address = {New York, NY, USA},
	series = {{CODASPY} '15},
	title = {{CoinParty}: {Secure} {Multi}-{Party} {Mixing} of {Bitcoins}},
	isbn = {978-1-4503-3191-3},
	shorttitle = {{CoinParty}},
	url = {https://dl.acm.org/doi/10.1145/2699026.2699100},
	doi = {10.1145/2699026.2699100},
	urldate = {2025-11-21},
	booktitle = {Proceedings of the 5th {ACM} {Conference} on {Data} and {Application} {Security} and {Privacy}},
	publisher = {Association for Computing Machinery},
	author = {Ziegeldorf, Jan Henrik and Grossmann, Fred and Henze, Martin and Inden, Nicolas and Wehrle, Klaus},
	month = mar,
	year = {2015},
	pages = {75--86},
}

@article{deuber_sok_2022,
	title = {{SoK}: {Assumptions} {Underlying} {Cryptocurrency} {Deanonymizations}},
	issn = {2299-0984},
	shorttitle = {{SoK}},
	url = {https://petsymposium.org/popets/2022/popets-2022-0091.php},
	urldate = {2025-11-21},
	journal = {Proceedings on Privacy Enhancing Technologies},
	author = {Deuber, Dominic and Ronge, Viktoria and Rückert, Christian},
	year = {2022},
}

@inproceedings{cannas2025jpeg,
  title={Is JPEG AI going to change image forensics?},
  author={Cannas, Edoardo Daniele and Mandelli, Sara and Popovic, Natasa and Alkhateeb, Ayman and Gnutti, Alessandro and Bestagini, Paolo and Tubaro, Stefano},
  booktitle={Proceedings of the IEEE/CVF International Conference on Computer Vision},
  pages={1564--1575},
  year={2025}
}

@article{gonzalez_arias_systematic_2024,
	title = {Systematic Review: Anti-Forensic Computer Techniques},
	volume = {14},
	rights = {http://creativecommons.org/licenses/by/3.0/},
	issn = {2076-3417},
	url = {https://www.mdpi.com/2076-3417/14/12/5302},
	doi = {10.3390/app14125302},
	shorttitle = {Systematic Review},
	pages = {5302},
	number = {12},
	journal = {Applied Sciences},
	author = {González Arias, Rafael and Bermejo Higuera, Javier and Rainer Granados, J. Javier and Bermejo Higuera, Juan Ramón and Sicilia Montalvo, Juan Antonio},
	urldate = {2025-03-06},
	date = {2024-01},
    year = {2024},
	langid = {english},
	note = {Number: 12
Publisher: Multidisciplinary Digital Publishing Institute},
}

@misc{the_grugq_art_2005,
	title = {The Art of Defiling - Defeating Forensic Analysis},
	url = {https://www.blackhat.com/presentations/bh-europe-05/bh-eu-05-grugq.pdf},
	note = {Blackhat Europe 2005},
	author = {the grugq},
	urldate = {2025-03-18},
	date = {2005},
}

@misc{liu_bleeding-edge_2006,
	title = {Bleeding-Edge Anti-Forensics},
	url = {https://resources.bishopfox.com/files/slides/2006/InfoSecWorld_2006-K2-Bleeding_Edge_AntiForensics-2006.pdf},
	note = {{InfoSec} World 2006},
	author = {Liu, Vincent and Brown, Francis},
	urldate = {2025-03-18},
	date = {2006},
}

@article{rogers_anti-forensics_2006,
	title = {Anti-Forensics: The Coming Wave in Digital Forensics},
	author = {Rogers, Marcus K},
	year = {2006},
	langid = {english},
}

@article{harris_arriving_2006,
	title = {Arriving at an anti-forensics consensus: Examining how to define and control the anti-forensics problem},
	volume = {3},
	issn = {1742-2876},
	url = {https://www.sciencedirect.com/science/article/pii/S1742287606000673},
	doi = {10.1016/j.diin.2006.06.005},
	series = {The Proceedings of the 6th Annual Digital Forensic Research Workshop ({DFRWS} '06)},
	shorttitle = {Arriving at an anti-forensics consensus},
	pages = {44--49},
	journal = {Digital Investigation},
	shortjournal = {Digital Investigation},
	author = {Harris, Ryan},
	urldate = {2025-03-18},
	date = {2006-09-01},
    year = {2006},
}

@article{kessler_anti-forensics_2007,
	title = {Anti-Forensics and the Digital Investigator},
	volume = {Edith Cowan University},
	url = {http://ro.ecu.edu.au/adf/1},
	doi = {10.4225/75/57AD39EE7FF25},
	pages = {December 3rd 2007},
	journal = {5th Australian Digital Forensics Conference},
	author = {Kessler, Gary},
	urldate = {2025-03-18},
	year = {2007},
	langid = {english},
	note = {Medium: {PDF}
Publisher: Security Research Institute ({SRI}), Edith Cowan University},
}

@inproceedings{sremack_taxonomy_2007,
	title = {Taxonomy of Anti-Computer Forensics Threats},
	volume = {P-114},
	booktitle = {Lecture Notes in Informatics ({LNI}), Proceedings - Series of the Gesellschaft fur Informatik ({GI})},
	pages = {103--112},
	author = {Sremack, J.C. and Antonov, A.V.},
	year = {2007},
}

@article{conlan_anti-forensics_2016,
	title = {Anti-forensics: Furthering digital forensic science through a new extended, granular taxonomy},
	volume = {18},
	issn = {17422876},
	url = {https://linkinghub.elsevier.com/retrieve/pii/S1742287616300378},
	doi = {10.1016/j.diin.2016.04.006},
	shorttitle = {Anti-forensics},
	pages = {S66--S75},
	journal = {Digital Investigation},
	shortjournal = {Digital Investigation},
	author = {Conlan, Kevin and Baggili, Ibrahim and Breitinger, Frank},
	urldate = {2025-02-12},
	date = {2016-08},
    year = {2016},
	langid = {english},
}

@misc{peron_digital_2010,
	title = {Digital Anti-Forensics: Emerging trends in data transformation techniques},
	url = {https://www.slideshare.net/slideshow/digital-antiforensics-emerging-trends-in-data-transformation-techniques/4050814},
	shorttitle = {Digital Anti-Forensics},
	publisher = {Seccuris Labs},
	author = {Peron, Christian S. J. and Legary, Michael},
	urldate = {2025-03-18},
	date = {2010-05-11},
	langid = {english},
}

@online{us_attorneys_office_southern_district_of_california_fbis_2021,
	title = {{FBI}’s Encrypted Phone Platform Infiltrated Hundreds of Criminal Syndicates; Result is Massive Worldwide Takedown},
	url = {https://www.justice.gov/usao-sdca/pr/fbi-s-encrypted-phone-platform-infiltrated-hundreds-criminal-syndicates-result-massive},
	author = {U.S. Attorney's Office, Southern District of California},
	urldate = {2025-04-08},
	date = {2021-06-08},
	langid = {english},
}

@article{Kitchenham_guidelines,
  author       = {Barbara Kitchenham and
                  Stuart Charters},
  title        = {Guidelines for performing Systematic Literature Reviews in Software Engineering Version 2.3},
  year         = {2007},
  url          = {https://www.researchgate.net/profile/Barbara-Kitchenham/publication/302924724_Guidelines_for_performing_Systematic_Literature_Reviews_in_Software_Engineering/links/61712932766c4a211c03a6f7/Guidelines-for-performing-Systematic-Literature-Reviews-in-Software-Engineering.pdf},
  institution = {School of Computer Science and Mathematics, Keele University and Department of Computer Science, University of Durham}
}

@article{Taneja_antiforensics_multimedia_review,
  author       = {Neeti Taneja and
                  Vijendra Singh Bramhe and
                  Dinesh Bhardwaj and
                  Ashu Taneja},
  title        = {Understanding digital image anti-forensics: an analytical review},
  journal      = {Multim. Tools Appl.},
  volume       = {83},
  number       = {4},
  pages        = {10445--10466},
  year         = {2024},
  url          = {https://doi.org/10.1007/s11042-023-15866-0},
  doi          = {10.1007/S11042-023-15866-0},
  bibsource    = {dblp computer science bibliography, https://dblp.org}
}

@misc{Interpol,
author = {Interpol},
title = {Digital forensics},
url = {https://www.interpol.int/How-we-work/Innovation/Digital-forensics}}

@inproceedings{adamu_conceptual_2020,
	title = {A {Conceptual} {Framework} for {Database} {Anti}-forensics {Impact} {Mitigation}},
	doi = {10.1109/ISDFS49300.2020.9116375},
	booktitle = {2020 8th {International} {Symposium} on {Digital} {Forensics} and {Security} ({ISDFS})},
	author = {Adamu, Bashir Zak and Karabatak, Murat and Ertam, Fatih},
	month = jun,
	year = {2020},
	pages = {1--6},
}

@inproceedings{ahmad_blockchain-based_2020,
	address = {New York, NY, USA},
	series = {{ARES} '20},
	title = {Blockchain-based chain of custody: towards real-time tamper-proof evidence management},
	isbn = {978-1-4503-8833-7},
	url = {https://doi.org/10.1145/3407023.3409199},
	doi = {10.1145/3407023.3409199},
	booktitle = {Proceedings of the 15th {International} {Conference} on {Availability}, {Reliability} and {Security}},
	publisher = {Association for Computing Machinery},
	author = {Ahmad, Liza and Khanji, Salam and Iqbal, Farkhund and Kamoun, Faouzi},
	year = {2020},
	note = {event-place: Virtual Event, Ireland},
}

@inproceedings{al-mousa_general_2021,
	title = {General {Countermeasures} of {Anti}-{Forensics} {Categories}},
	doi = {10.1109/GC-ElecEng52322.2021.9788230},
	booktitle = {2021 {Global} {Congress} on {Electrical} {Engineering} ({GC}-{ElecEng})},
	author = {Al-Mousa, Mohammad Rasmi and Sweerky, Nael A. and Samara, Ghassan and Alghanim, Mohammed and Hussein, Abla Suleiman Ismail and Qadoumi, Braa},
	month = dec,
	year = {2021},
	pages = {5--10},
}

@inproceedings{al_fahdi_challenges_2013,
	title = {Challenges to digital forensics: {A} survey of researchers \& practitioners attitudes and opinions},
	doi = {10.1109/ISSA.2013.6641058},
	booktitle = {2013 {Information} {Security} for {South} {Africa}},
	author = {Al Fahdi, M. and Clarke, N.L. and Furnell, S.M.},
	month = aug,
	year = {2013},
	note = {ISSN: 2330-9881},
	pages = {1--8},
}

@article{alfosail_tor_2021,
	title = {Tor forensics: {Proposed} workflow for client memory artefacts},
	volume = {106},
	issn = {0167-4048},
	url = {https://www.sciencedirect.com/science/article/pii/S0167404821001358},
	doi = {https://doi.org/10.1016/j.cose.2021.102311},
	journal = {Computers \& Security},
	author = {Alfosail, Malak and Norris, Peter},
	year = {2021},
	pages = {102311},
}

@article{bakas_detection_2021,
	title = {Detection and localization of inter-frame forgeries in videos based on macroblock variation and motion vector analysis},
	volume = {89},
	issn = {0045-7906},
	url = {https://www.sciencedirect.com/science/article/pii/S0045790620307783},
	doi = {https://doi.org/10.1016/j.compeleceng.2020.106929},
	journal = {Computers \& Electrical Engineering},
	author = {Bakas, Jamimamul and Naskar, Ruchira and Bakshi, Sambit},
	year = {2021},
	pages = {106929},
}

@inproceedings{bellekens_data_2015,
	title = {Data remanence and digital forensic investigation for {CUDA} {Graphics} {Processing} {Units}},
	doi = {10.1109/INM.2015.7140493},
	booktitle = {2015 {IFIP}/{IEEE} {International} {Symposium} on {Integrated} {Network} {Management} ({IM})},
	author = {Bellekens, Xavier and Paul, Greig and Irvine, James M. and Tachtatzis, Christos and Atkinson, Robert C. and Kirkham, Tony and Renfrew, Craig},
	month = may,
	year = {2015},
	note = {ISSN: 1573-0077},
	pages = {1345--1350},
}

@inproceedings{botas_counterfeiting_2015,
	title = {Counterfeiting and {Defending} the {Digital} {Forensic} {Process}},
	doi = {10.1109/CIT/IUCC/DASC/PICOM.2015.291},
	booktitle = {2015 {IEEE} {International} {Conference} on {Computer} and {Information} {Technology}; {Ubiquitous} {Computing} and {Communications}; {Dependable}, {Autonomic} and {Secure} {Computing}; {Pervasive} {Intelligence} and {Computing}},
	author = {Botas, Alvaro and Rodriguez, Ricardo J. and Väisänen, Teemu and Zdzichowski, Patrycjusz},
	month = oct,
	year = {2015},
	pages = {1966--1971},
}

@article{carrier_different_2010,
	title = {Different interpretations of {ISO9660} file systems},
	volume = {7},
	issn = {1742-2876},
	url = {https://www.sciencedirect.com/science/article/pii/S1742287610000435},
	doi = {https://doi.org/10.1016/j.diin.2010.05.016},
	journal = {Digital Investigation},
	author = {Carrier, Brian D.},
	year = {2010},
	pages = {S129--S134},
}

@article{casey_growing_2011,
	title = {The growing impact of full disk encryption on digital forensics},
	volume = {8},
	issn = {1742-2876},
	url = {https://www.sciencedirect.com/science/article/pii/S1742287611000727},
	doi = {https://doi.org/10.1016/j.diin.2011.09.005},
	number = {2},
	journal = {Digital Investigation},
	author = {Casey, Eoghan and Fellows, Geoff and Geiger, Matthew and Stellatos, Gerasimos},
	year = {2011},
	pages = {129--134},
}

@article{castiglione_automated_2013,
	title = {Automated {Production} of {Predetermined} {Digital} {Evidence}},
	volume = {1},
	issn = {2169-3536},
	doi = {10.1109/ACCESS.2013.2260817},
	journal = {IEEE Access},
	author = {Castiglione, Aniello and Cattaneo, Giuseppe and De Maio, Giancarlo and De Santis, Alfredo},
	year = {2013},
	pages = {216--231},
}

@inproceedings{castiglione_forensic_2012,
	title = {The {Forensic} {Analysis} of a {False} {Digital} {Alibi}},
	doi = {10.1109/IMIS.2012.127},
	booktitle = {2012 {Sixth} {International} {Conference} on {Innovative} {Mobile} and {Internet} {Services} in {Ubiquitous} {Computing}},
	author = {Castiglione, Aniello and Cattaneo, Giuseppe and De Maio, Giancarlo and De Santis, Alfredo and Costabile, Gerardo and Epifani, Mattia},
	month = jul,
	year = {2012},
	pages = {114--121},
}

@article{castiglione_taking_2007,
	title = {Taking advantages of a disadvantage: {Digital} forensics and steganography using document metadata},
	volume = {80},
	issn = {0164-1212},
	url = {https://www.sciencedirect.com/science/article/pii/S0164121206001981},
	doi = {https://doi.org/10.1016/j.jss.2006.07.006},
	number = {5},
	journal = {Journal of Systems and Software},
	author = {Castiglione, A. and Santis, A. De and Soriente, C.},
	year = {2007},
	pages = {750--764},
}

@inproceedings{dafale_sensor_2017,
	title = {Sensor pattern noise based source anonymization},
	doi = {10.1109/SSPS.2017.8071572},
	booktitle = {2017 {Third} {International} {Conference} on {Sensing}, {Signal} {Processing} and {Security} ({ICSSS})},
	author = {Dafale, Ninad N. and Naskar, Ruchira},
	month = may,
	year = {2017},
	pages = {93--98},
}

@inproceedings{dahbur_anti-forensics_2011,
	address = {New York, NY, USA},
	series = {{ISWSA} '11},
	title = {The anti-forensics challenge},
	isbn = {978-1-4503-0474-0},
	url = {https://doi.org/10.1145/1980822.1980836},
	doi = {10.1145/1980822.1980836},
	booktitle = {Proceedings of the 2011 {International} {Conference} on {Intelligent} {Semantic} {Web}-{Services} and {Applications}},
	publisher = {Association for Computing Machinery},
	author = {Dahbur, Kamal and Mohammad, Bassil},
	year = {2011},
	note = {event-place: Amman, Jordan},
}

@inproceedings{franzen_randcompile_2023,
	address = {New York, NY, USA},
	series = {{ACSAC} '23},
	title = {{RandCompile}: {Removing} {Forensic} {Gadgets} from the {Linux} {Kernel} to {Combat} its {Analysis}},
	isbn = {979-8-4007-0886-2},
	url = {https://doi.org/10.1145/3627106.3627197},
	doi = {10.1145/3627106.3627197},
	booktitle = {Proceedings of the 39th {Annual} {Computer} {Security} {Applications} {Conference}},
	publisher = {Association for Computing Machinery},
	author = {Franzen, Fabian and Wilhelmer, Andreas Chris and Grossklags, Jens},
	year = {2023},
	note = {event-place: Austin, TX, USA},
	pages = {677--690},
}

@article{freiling_controlled_2018,
	title = {Controlled experiments in digital evidence tampering},
	volume = {24},
	issn = {1742-2876},
	url = {https://www.sciencedirect.com/science/article/pii/S1742287618300434},
	doi = {https://doi.org/10.1016/j.diin.2018.01.011},
	journal = {Digital Investigation},
	author = {Freiling, Felix and Hösch, Leonhard},
	year = {2018},
	pages = {S83--S92},
}

@inproceedings{galhuber_time_2021,
	address = {New York, NY, USA},
	series = {{ARES} '21},
	title = {Time for {Truth}: {Forensic} {Analysis} of {NTFS} {Timestamps}},
	isbn = {978-1-4503-9051-4},
	url = {https://doi.org/10.1145/3465481.3470016},
	doi = {10.1145/3465481.3470016},
	booktitle = {Proceedings of the 16th {International} {Conference} on {Availability}, {Reliability} and {Security}},
	publisher = {Association for Computing Machinery},
	author = {Galhuber, Michael and Luh, Robert},
	year = {2021},
	note = {event-place: Vienna, Austria},
}

@article{gobel_anti-forensics_2018,
	title = {Anti-forensics in ext4: {On} secrecy and usability of timestamp-based data hiding},
	volume = {24},
	issn = {1742-2876},
	url = {https://www.sciencedirect.com/science/article/pii/S174228761830046X},
	doi = {https://doi.org/10.1016/j.diin.2018.01.014},
	journal = {Digital Investigation},
	author = {Göbel, Thomas and Baier, Harald},
	year = {2018},
	pages = {S111--S120},
}

@inproceedings{gobel_revisiting_2019,
	address = {New York, NY, USA},
	series = {{ARES} '19},
	title = {Revisiting {Data} {Hiding} {Techniques} for {Apple} {File} {System}},
	isbn = {978-1-4503-7164-3},
	url = {https://doi.org/10.1145/3339252.3340524},
	doi = {10.1145/3339252.3340524},
	booktitle = {Proceedings of the 14th {International} {Conference} on {Availability}, {Reliability} and {Security}},
	publisher = {Association for Computing Machinery},
	author = {Göbel, Thomas and Türr, Jan and Baier, Harald},
	year = {2019},
	note = {event-place: Canterbury, CA, United Kingdom},
}

@article{goh_plausibly-deniable_2011,
	title = {A {Plausibly}-{Deniable}, {Practical} {Trusted} {Platform} {Module} {Based} {Anti}-{Forensics} {Client}-{Server} {System}},
	volume = {29},
	issn = {1558-0008},
	doi = {10.1109/JSAC.2011.110805},
	number = {7},
	journal = {IEEE Journal on Selected Areas in Communications},
	author = {Goh, Weihan and Leong, Peng Chor and Yeo, Chai Kiat},
	month = aug,
	year = {2011},
	pages = {1377--1391},
}

@inproceedings{gul_survey_2017,
	title = {A survey on anti-forensics techniques},
	doi = {10.1109/IDAP.2017.8090341},
	booktitle = {2017 {International} {Artificial} {Intelligence} and {Data} {Processing} {Symposium} ({IDAP})},
	author = {Gül, Murat and Kugu, Emin},
	month = sep,
	year = {2017},
	pages = {1--6},
}

@article{hasanabadi_game-theoretic_2020,
	title = {A game-theoretic defensive approach for forensic investigators against rootkits},
	volume = {33},
	issn = {2666-2817},
	url = {https://www.sciencedirect.com/science/article/pii/S2666281720300299},
	doi = {https://doi.org/10.1016/j.fsidi.2020.200909},
	journal = {Forensic Science International: Digital Investigation},
	author = {Hasanabadi, Saeed Shafiee and Lashkari, Arash Habibi and Ghorbani, Ali A.},
	year = {2020},
	pages = {200909},
}

@article{hasanabadi_memory-based_2021,
	title = {A memory-based game-theoretic defensive approach for digital forensic investigators},
	volume = {38},
	issn = {2666-2817},
	url = {https://www.sciencedirect.com/science/article/pii/S2666281721001220},
	doi = {https://doi.org/10.1016/j.fsidi.2021.301214},
	journal = {Forensic Science International: Digital Investigation},
	author = {Hasanabadi, Saeed Shafiee and Lashkari, Arash Habibi and Ghorbani, Ali A.},
	year = {2021},
	pages = {301214},
}

@article{hasanabadi_survey_2020,
	title = {A survey and research challenges of anti-forensics: {Evaluation} of game-theoretic models in simulation of forensic agents’ behaviour},
	volume = {35},
	issn = {2666-2817},
	url = {https://www.sciencedirect.com/science/article/pii/S2666281720300925},
	doi = {https://doi.org/10.1016/j.fsidi.2020.301024},
	journal = {Forensic Science International: Digital Investigation},
	author = {Hasanabadi, Saeed Shafiee and Lashkari, Arash Habibi and Ghorbani, Ali A.},
	year = {2020},
	pages = {301024},
}

@inproceedings{hausknecht_anti-computer_2017,
	title = {Anti-computer forensics},
	doi = {10.23919/MIPRO.2017.7973612},
	booktitle = {2017 40th {International} {Convention} on {Information} and {Communication} {Technology}, {Electronics} and {Microelectronics} ({MIPRO})},
	author = {Hausknecht, K. and Gruičić, S.},
	month = may,
	year = {2017},
	pages = {1233--1240},
}

@inproceedings{heeger_exhide_2021,
	address = {New York, NY, USA},
	series = {{ARES} '21},
	title = {{exHide}: {Hiding} {Data} within the {exFAT} {File} {System}},
	isbn = {978-1-4503-9051-4},
	url = {https://doi.org/10.1145/3465481.3470117},
	doi = {10.1145/3465481.3470117},
	booktitle = {Proceedings of the 16th {International} {Conference} on {Availability}, {Reliability} and {Security}},
	publisher = {Association for Computing Machinery},
	author = {Heeger, Julian and Yannikos, York and Steinebach, Martin},
	year = {2021},
	note = {event-place: Vienna, Austria},
}

@inproceedings{hermon_forensic_2022,
	title = {Forensic {Techniques} to {Detect} {Hidden} {Data} in {Alternate} {Data} {Streams} in {NTFS}},
	doi = {10.1109/IBSSC56953.2022.10037507},
	booktitle = {2022 {IEEE} {Bombay} {Section} {Signature} {Conference} ({IBSSC})},
	author = {Hermon, Rahul and Singh, Upasna and Singh, Bhupendra},
	month = dec,
	year = {2022},
	pages = {1--6},
}

@inproceedings{hsu_digital_2011,
	title = {A digital evidence protection method with hierarchical access control mechanisms},
	doi = {10.1109/CCST.2011.6095882},
	booktitle = {2011 {Carnahan} {Conference} on {Security} {Technology}},
	author = {Hsu, Chien-Lung and Liu, Boo-Chen and Lin, Yu-Li},
	month = oct,
	year = {2011},
	note = {ISSN: 2153-0742},
	pages = {1--9},
}

@article{huebner_data_2006,
	title = {Data hiding in the {NTFS} file system},
	volume = {3},
	issn = {1742-2876},
	url = {https://www.sciencedirect.com/science/article/pii/S1742287606001265},
	doi = {https://doi.org/10.1016/j.diin.2006.10.005},
	number = {4},
	journal = {Digital Investigation},
	author = {Huebner, Ewa and Bem, Derek and Wee, Cheong Kai},
	year = {2006},
	pages = {211--226},
}

@inproceedings{jain_anti-forensics_2014,
	title = {Anti-forensics techniques: {An} analytical review},
	doi = {10.1109/IC3.2014.6897209},
	booktitle = {2014 {Seventh} {International} {Conference} on {Contemporary} {Computing} ({IC3})},
	author = {Jain, Anu and Chhabra, Gurpal Singh},
	month = aug,
	year = {2014},
	pages = {412--418},
}

@inproceedings{jang_understanding_2016,
	title = {Understanding {Anti}-forensic {Techniques} with {Timestamp} {Manipulation} ({Invited} {Paper})},
	doi = {10.1109/IRI.2016.94},
	booktitle = {2016 {IEEE} 17th {International} {Conference} on {Information} {Reuse} and {Integration} ({IRI})},
	author = {Jang, Dae-il and Ahn, Gail-Joon and Hwang, Hyunuk and Kim, Kibom},
	month = jul,
	year = {2016},
	pages = {609--614},
}

@article{jones_what_2004,
	title = {What evidence is left after disk cleaners?},
	volume = {1},
	issn = {1742-2876},
	url = {https://www.sciencedirect.com/science/article/pii/S1742287604000568},
	doi = {https://doi.org/10.1016/j.diin.2004.07.002},
	number = {3},
	journal = {Digital Investigation},
	author = {Jones, Andy and Meyler, Christopher},
	year = {2004},
	pages = {183--188},
}

@article{karampidis_review_2018,
	title = {A review of image steganalysis techniques for digital forensics},
	volume = {40},
	issn = {2214-2126},
	url = {https://www.sciencedirect.com/science/article/pii/S2214212617300777},
	doi = {https://doi.org/10.1016/j.jisa.2018.04.005},
	journal = {Journal of Information Security and Applications},
	author = {Karampidis, Konstantinos and Kavallieratou, Ergina and Papadourakis, Giorgos},
	year = {2018},
	pages = {217--235},
}

@article{khan_designing_2011,
	title = {Designing a cluster-based covert channel to evade disk investigation and forensics},
	volume = {30},
	issn = {0167-4048},
	url = {https://www.sciencedirect.com/science/article/pii/S016740481000088X},
	doi = {https://doi.org/10.1016/j.cose.2010.10.005},
	number = {1},
	journal = {Computers \& Security},
	author = {Khan, Hassan and Javed, Mobin and Khayam, Syed Ali and Mirza, Fauzan},
	year = {2011},
	pages = {35--49},
}

@inproceedings{kim_null_2022,
	address = {New York, NY, USA},
	series = {{MobiHoc} '22},
	title = {{NULL} byte injection: anti-forensic technique for data hiding in {FAT32} file system},
	isbn = {978-1-4503-9165-8},
	url = {https://doi.org/10.1145/3492866.3558587},
	doi = {10.1145/3492866.3558587},
	booktitle = {Proceedings of the {Twenty}-{Third} {International} {Symposium} on {Theory}, {Algorithmic} {Foundations}, and {Protocol} {Design} for {Mobile} {Networks} and {Mobile} {Computing}},
	publisher = {Association for Computing Machinery},
	author = {Kim, Donghyun and Lee, Youn Kyu and Jeong, Jongwook},
	year = {2022},
	note = {event-place: Seoul, Republic of Korea},
	pages = {265--270},
}

@inproceedings{kwon_tool_2008,
	title = {A {Tool} for the {Detection} of {Hidden} {Data} in {Microsoft} {Compound} {Document} {File} {Format}},
	doi = {10.1109/ICISS.2008.19},
	booktitle = {2008 {International} {Conference} on {Information} {Science} and {Security} ({ICISS} 2008)},
	author = {Kwon, Hyukdon and Kim, Yeog and Lee, Sangjin and Lim, Jongin},
	month = jan,
	year = {2008},
	pages = {141--146},
}

@inproceedings{liu_using_2012,
	title = {Using {Attack} {Graphs} in {Forensic} {Examinations}},
	doi = {10.1109/ARES.2012.58},
	booktitle = {2012 {Seventh} {International} {Conference} on {Availability}, {Reliability} and {Security}},
	author = {Liu, Changwei and Singhal, Anoop and Wijesekera, Duminda},
	month = aug,
	year = {2012},
	pages = {596--603},
}

@inproceedings{lowetz_anti-forensics_2024,
	title = {Anti-forensics {Under} {Scrutiny} {Assessing} the {Effectiveness} of {Digital} {Obfuscation} in the {Cloud}},
	doi = {10.1109/ISDFS60797.2024.10526470},
	booktitle = {2024 12th {International} {Symposium} on {Digital} {Forensics} and {Security} ({ISDFS})},
	author = {Lowetz, Christopher and Shepard, Grant and Coffman, Joel},
	month = apr,
	year = {2024},
	note = {ISSN: 2768-1831},
	pages = {1--6},
}

@inproceedings{martin_data_2022,
	address = {New York, NY, USA},
	series = {{ICFNDS} '21},
	title = {Data {Preservation} {System} using {BoCA}: {Blockchain}-of-{Custody} {Application}},
	isbn = {978-1-4503-8734-7},
	url = {https://doi.org/10.1145/3508072.3508084},
	doi = {10.1145/3508072.3508084},
	booktitle = {Proceedings of the 5th {International} {Conference} on {Future} {Networks} and {Distributed} {Systems}},
	publisher = {Association for Computing Machinery},
	author = {Martin, Thomas and Hammoudeh, Mohammad},
	year = {2022},
	note = {event-place: Dubai, United Arab Emirates},
	pages = {70--77},
}

@inproceedings{martini_detecting_2008,
	title = {Detecting and {Manipulating} {Compressed} {Alternate} {Data} {Streams} in a {Forensics} {Investigation}},
	doi = {10.1109/WDFIA.2008.9},
	booktitle = {2008 {Third} {International} {Annual} {Workshop} on {Digital} {Forensics} and {Incident} {Analysis}},
	author = {Martini, Adamantini I. and Zaharis, Alexandros and Ilioudis, Christos},
	month = oct,
	year = {2008},
	pages = {53--59},
}

@inproceedings{mcdonald_enhanced_2017,
	title = {Enhanced {Operating} {System} {Protection} to {Support} {Digital} {Forensic} {Investigations}},
	doi = {10.1109/Trustcom/BigDataSE/ICESS.2017.296},
	booktitle = {2017 {IEEE} {Trustcom}/{BigDataSE}/{ICESS}},
	author = {McDonald, Jeffrey Todd and Manikyam, Ramya and Glisson, William Bradley and Andel, Todd R. and Gu, Yuan Xiang},
	month = aug,
	year = {2017},
	note = {ISSN: 2324-9013},
	pages = {650--659},
}

@article{mohamed_detection_2019,
	title = {Detection of {Timestamps} {Tampering} in {NTFS} using {Machine} {Learning}},
	volume = {160},
	issn = {1877-0509},
	url = {https://www.sciencedirect.com/science/article/pii/S1877050919317119},
	doi = {https://doi.org/10.1016/j.procs.2019.11.011},
	journal = {Procedia Computer Science},
	author = {Mohamed, Alji and Khalid, Chougdali},
	year = {2019},
	pages = {778--784},
}

@article{neale_case_2022,
	title = {The case for {Zero} {Trust} {Digital} {Forensics}},
	volume = {40},
	issn = {2666-2817},
	url = {https://www.sciencedirect.com/science/article/pii/S266628172200021X},
	doi = {https://doi.org/10.1016/j.fsidi.2022.301352},
	journal = {Forensic Science International: Digital Investigation},
	author = {Neale, Christopher and Kennedy, Ian and Price, Blaine and Yu, Yijun and Nuseibeh, Bashar},
	year = {2022},
	pages = {301352},
}

@article{neale_fool_2023,
	title = {Fool me once: {A} systematic review of techniques to authenticate digital artefacts},
	volume = {45},
	issn = {2666-2817},
	shorttitle = {Fool me once},
	url = {https://www.sciencedirect.com/science/article/pii/S2666281723000173},
	doi = {10.1016/j.fsidi.2023.301516},
	language = {en},
	urldate = {2023-02-23},
	journal = {Forensic Science International: Digital Investigation},
	author = {Neale, Christopher},
	month = jun,
	year = {2023},
	pages = {301516},
}

@article{nowroozi_employing_2023,
	title = {Employing {Deep} {Ensemble} {Learning} for {Improving} the {Security} of {Computer} {Networks} {Against} {Adversarial} {Attacks}},
	volume = {20},
	issn = {1932-4537},
	doi = {10.1109/TNSM.2023.3267831},
	number = {2},
	journal = {IEEE Transactions on Network and Service Management},
	author = {Nowroozi, Ehsan and Mohammadi, Mohammadreza and Savaş, Erkay and Mekdad, Yassine and Conti, Mauro},
	month = jun,
	year = {2023},
	pages = {2096--2105},
}

@inproceedings{odebade_mitigating_2017,
	title = {Mitigating anti-forensics in the {Cloud} via resource-based privacy preserving activity attribution},
	url = {https://ieeexplore.ieee.org/abstract/document/7939155},
	doi = {10.1109/SDS.2017.7939155},
	urldate = {2024-10-11},
	booktitle = {2017 {Fourth} {International} {Conference} on {Software} {Defined} {Systems} ({SDS})},
	author = {Odebade, Adeyinka and Welsh, Thomas and Mthunzi, Siyakha and Benkhelifa, Elhadj},
	month = may,
	year = {2017},
	pages = {143--149},
}

@article{oh_forensic_2024,
	title = {Forensic {Detection} of {Timestamp} {Manipulation} for {Digital} {Forensic} {Investigation}},
	volume = {12},
	issn = {2169-3536},
	doi = {10.1109/ACCESS.2024.3395644},
	journal = {IEEE Access},
	author = {Oh, Junghoon and Lee, Sangjin and Hwang, Hyunuk},
	year = {2024},
	pages = {72544--72565},
}

@article{palutke_styx_2018,
	title = {Styx: {Countering} robust memory acquisition},
	volume = {24},
	issn = {1742-2876},
	url = {https://www.sciencedirect.com/science/article/pii/S1742287618300367},
	doi = {https://doi.org/10.1016/j.diin.2018.01.004},
	journal = {Digital Investigation},
	author = {Palutke, Ralph and Freiling, Felix},
	year = {2018},
	pages = {S18--S28},
}

@inproceedings{raj_counter_2019,
	title = {Counter {Forensics}: {A} {New} {PRNU} {Based} {Method} for {Image} {Source} {Anonymization}},
	doi = {10.1109/ICECCT.2019.8868948},
	booktitle = {2019 {IEEE} {International} {Conference} on {Electrical}, {Computer} and {Communication} {Technologies} ({ICECCT})},
	author = {Raj, Aiswariya and Sankar, Deepa},
	month = feb,
	year = {2019},
	pages = {1--7},
}

@article{rekhis_hierarchical_2012,
	title = {A {Hierarchical} {Visibility} theory for formal digital investigation of anti-forensic attacks},
	volume = {31},
	issn = {0167-4048},
	url = {https://www.sciencedirect.com/science/article/pii/S0167404812001022},
	doi = {https://doi.org/10.1016/j.cose.2012.06.009},
	number = {8},
	journal = {Computers \& Security},
	author = {Rekhis, Slim and Boudriga, Noureddine},
	year = {2012},
	pages = {967--982},
}

@inproceedings{savoldi_data_2007,
	title = {Data {Hiding} in {SIM}/{USIM} {Cards}: {A} {Steganographic} {Approach}},
	doi = {10.1109/SADFE.2007.7},
	booktitle = {Second {International} {Workshop} on {Systematic} {Approaches} to {Digital} {Forensic} {Engineering} ({SADFE}'07)},
	author = {Savoldi, Antonio and Gubian, Paolo},
	month = apr,
	year = {2007},
	pages = {86--100},
}

@article{savoldi_statistical_2012,
	title = {A statistical method for detecting on-disk wiped areas},
	volume = {8},
	issn = {1742-2876},
	url = {https://www.sciencedirect.com/science/article/pii/S1742287611000545},
	doi = {https://doi.org/10.1016/j.diin.2011.06.005},
	number = {3},
	journal = {Digital Investigation},
	author = {Savoldi, Antonio and Piccinelli, Mario and Gubian, Paolo},
	year = {2012},
	pages = {194--214},
}

@inproceedings{schmitt_introducing_2018,
	title = {Introducing {Anti}-{Forensics} to {SQLite} {Corpora} and {Tool} {Testing}},
	doi = {10.1109/IMF.2018.00014},
	booktitle = {2018 11th {International} {Conference} on {IT} {Security} {Incident} {Management} \& {IT} {Forensics} ({IMF})},
	author = {Schmitt, Sven},
	month = may,
	year = {2018},
	pages = {89--106},
}

@article{schneider_ambiguous_2022,
	title = {Ambiguous file system partitions},
	volume = {42},
	issn = {2666-2817},
	url = {https://www.sciencedirect.com/science/article/pii/S2666281722000804},
	doi = {https://doi.org/10.1016/j.fsidi.2022.301399},
	journal = {Forensic Science International: Digital Investigation},
	author = {Schneider, Janine and Eichhorn, Maximilian and Freiling, Felix},
	year = {2022},
	pages = {301399},
}

@article{schneider_prudent_2022,
	title = {Prudent design principles for digital tampering experiments},
	volume = {40},
	issn = {2666-2817},
	url = {https://www.sciencedirect.com/science/article/pii/S2666281722000038},
	doi = {https://doi.org/10.1016/j.fsidi.2022.301334},
	journal = {Forensic Science International: Digital Investigation},
	author = {Schneider, Janine and Düsel, Linus and Lorch, Benedikt and Drafz, Julia and Freiling, Felix},
	year = {2022},
	pages = {301334},
}

@article{schneider_tampering_2020,
	title = {Tampering with {Digital} {Evidence} is {Hard}: {The} {Case} of {Main} {Memory} {Images}},
	volume = {32},
	issn = {2666-2817},
	url = {https://www.sciencedirect.com/science/article/pii/S2666281720300196},
	doi = {https://doi.org/10.1016/j.fsidi.2020.300924},
	journal = {Forensic Science International: Digital Investigation},
	author = {Schneider, Janine and Wolf, Julian and Freiling, Felix},
	year = {2020},
	pages = {300924},
}

@inproceedings{shahaab_preventing_2021,
	address = {New York, NY, USA},
	series = {{ICBCT} '21},
	title = {Preventing {Spoliation} of {Evidence} with {Blockchain}: {A} {Perspective} from {South} {Asia}},
	isbn = {978-1-4503-8962-4},
	url = {https://doi.org/10.1145/3460537.3460550},
	doi = {10.1145/3460537.3460550},
	booktitle = {Proceedings of the 2021 3rd {International} {Conference} on {Blockchain} {Technology}},
	publisher = {Association for Computing Machinery},
	author = {Shahaab, Ali and Hewage, Chaminda and Khan, Imtiaz},
	year = {2021},
	note = {event-place: Shanghai, China},
	pages = {45--52},
}

@article{valente_improving_2019,
	title = {Improving the {Security} of {Visual} {Challenges}},
	volume = {3},
	issn = {2378-962X},
	url = {https://doi.org/10.1145/3331183},
	doi = {10.1145/3331183},
	number = {3},
	journal = {ACM Trans. Cyber-Phys. Syst.},
	author = {Valente, Junia and Bahirat, Kanchan and Venechanos, Kelly and Cardenas, Alvaro A. and Balakrishnan, Prabhakaran},
	month = aug,
	year = {2019},
	note = {Place: New York, NY, USA
Publisher: Association for Computing Machinery},
}

@inproceedings{wundram_anti-forensics_2013,
	title = {Anti-forensics: {The} {Next} {Step} in {Digital} {Forensics} {Tool} {Testing}},
	doi = {10.1109/IMF.2013.17},
	booktitle = {2013 {Seventh} {International} {Conference} on {IT} {Security} {Incident} {Management} and {IT} {Forensics}},
	author = {Wundram, Martin and Freiling, Felix C. and Moch, Christian},
	month = mar,
	year = {2013},
	pages = {83--97},
}

@article{yu_covert_2015,
	title = {Covert communication by means of email spam: {A} challenge for digital investigation},
	volume = {13},
	issn = {1742-2876},
	url = {https://www.sciencedirect.com/science/article/pii/S1742287615000432},
	doi = {https://doi.org/10.1016/j.diin.2015.04.003},
	journal = {Digital Investigation},
	author = {Yu, Szde},
	year = {2015},
	pages = {72--79},
}

@article{zax_faust_2009,
	title = {{FAUST}: {Forensic} artifacts of uninstalled steganography tools},
	volume = {6},
	issn = {1742-2876},
	url = {https://www.sciencedirect.com/science/article/pii/S1742287609000267},
	doi = {https://doi.org/10.1016/j.diin.2009.02.002},
	number = {1},
	journal = {Digital Investigation},
	author = {Zax, Rachel and Adelstein, Frank},
	year = {2009},
	pages = {25--38},
}

@inproceedings{zubair_control_2022,
	title = {Control {Logic} {Obfuscation} {Attack} in {Industrial} {Control} {Systems}},
	doi = {10.1109/CSR54599.2022.9850326},
	booktitle = {2022 {IEEE} {International} {Conference} on {Cyber} {Security} and {Resilience} ({CSR})},
	author = {Zubair, Nauman and Ayub, Adeen and Yoo, Hyunguk and Ahmed, Irfan},
	month = jul,
	year = {2022},
	pages = {227--232},
}

\clearpage
\onecolumn
\section*{Appendix}
\vspace*{\fill}
\begin{table}[htbp]
    \centering
    \scriptsize
    \caption{Overview of included papers with type anti-forensic and their subdomains, methodology, and targets.}
    \label{tbl:overview:1}
    \begin{threeparttable}
        % \addtolength{\tabcolsep}{-0.75em}
%
    \end{threeparttable}
    \begin{tablenotes}
        \tiny
        \item Legend: \; \circfull{} Yes, \circempty{} No, \circhalf{} Software/Dataset was developed but is not available
    \end{tablenotes}
\end{table}
\vspace*{\fill}

\begin{table*}[htbp]
    \centering
    \scriptsize
    \caption{Overview of included papers with type counter-anti-forensic and their subdomains, methodology, and targets.}
    \label{tbl:overview:2}
    \begin{threeparttable}
        % \addtolength{\tabcolsep}{-0.75em}
% %
    \end{threeparttable}
    \begin{tablenotes}
        \tiny
        \item Legend: \; \circfull{} Yes, \circempty{} No, \circhalf{} Software/Dataset was developed but is not available
    \end{tablenotes}
\end{table*}

\begin{table*}[htbp]
    \centering
    \scriptsize
    \caption{Overview of included papers with type both and their subdomains, methodology, and targets.}
    \label{tbl:overview:3}
    \begin{threeparttable}
        % \addtolength{\tabcolsep}{-0.75em}
% %
    \end{threeparttable}
    \begin{tablenotes}
        \tiny
        \item Legend: \; \circfull{} Yes, \circempty{} No, \circhalf{} Software/Dataset was developed but is not available
    \end{tablenotes}
\end{table*}

\end{document}